\documentclass[twocolumn]{aastex63}
\usepackage{amsmath}
\usepackage{mathrsfs}
\usepackage{xcolor}
\usepackage[T1]{fontenc}
\usepackage{fourier}
\usepackage{hyperref}
\usepackage[normalem]{ulem}
\usepackage{relsize}

\newcommand{\Mej}{$M_\text{ej}$} 
\newcommand{\Msolar}{$M_\Sun$} 
 
\newcommand{\Mni}{$M_{\rm Ni}$}

\newcommand{\nickel}{$^{56}$Ni }
\newcommand{\nickelwospace}{$^{56}$Ni}

\newcommand{\nilou}{\color{black}}
\newcommand{\nil}{}
\received{}
\revised{}
\accepted{}
\submitjournal{ApJ}

\shorttitle{NickelofStrippedEnvelopeSNe}
\shortauthors{Afsariardchi et al.}

\begin{document}

\title{The Nickel Mass Distribution of Stripped-Envelope Supernovae: Implications for Additional Power Sources}

\correspondingauthor{Niloufar Afsariardchi}
\email{afsariardchi@astro.utoronto.ca}

\author[0000-0002-1338-490X]{Niloufar Afsariardchi}
\affiliation{David A. Dunlap Department of Astronomy and Astrophysics, University of Toronto,\\ 50 St. George Street, Toronto, Ontario, M5S 3H4 Canada}

\author[0000-0001-7081-0082]{Maria R. Drout}
\affiliation{David A. Dunlap Department of Astronomy and Astrophysics, University of Toronto,\\ 50 St. George Street, Toronto, Ontario, M5S 3H4 Canada}
\affiliation{The Observatories of the Carnegie Institution for Science, 813 Santa Barbara St, Pasadena, CA, 91101, USA}

\author[0000-0003-4307-0589]{David K. Khatami}
\affiliation{Department of Astronomy, University of California, Berkeley, CA, 94720}

\author[0000-0001-9732-2281]{Christopher D. Matzner}
\affiliation{David A. Dunlap Department of Astronomy and Astrophysics, University of Toronto,\\ 50 St. George Street, Toronto, Ontario, M5S 3H4 Canada}

\author[0000-0003-4200-5064]{Dae-Sik Moon}
\affiliation{David A. Dunlap Department of Astronomy and Astrophysics, University of Toronto,\\ 50 St. George Street, Toronto, Ontario, M5S 3H4 Canada}

\author[0000-0003-3656-5268]{Yuan Qi Ni}
\affiliation{David A. Dunlap Department of Astronomy and Astrophysics, University of Toronto,\\ 50 St. George Street, Toronto, Ontario, M5S 3H4 Canada}



\begin{abstract}
We perform a systematic study of the \nickel mass (\Mni) of 27 stripped envelope supernovae (SESNe) by modeling their light-curve tails, highlighting that use of ``Arnett's rule'' overestimates \Mni\ for SESN by a factor of $\sim$2. Recently, \citet{Khatami2019} presented a new model relating the peak time ($t_{\rm p}$) and luminosity ($L_{\rm p}$) of a radioactive-powered SN to its \Mni\ that addresses several limitations of Arnett-like models, but depends on a dimensionless parameter, $\beta$. Using observed $t_{\rm p}$, $L_{\rm p}$, and tail-measured \Mni\ values for 27 SESN, we observationally calibrate $\beta$ for the first time.  Despite scatter, we demonstrate that the model of \citet{Khatami2019} with empirically-calibrated $\beta$ values provides significantly improved measurements of \Mni\ when only photospheric data is available.  However, these observationally-constrained $\beta$ values are systematically lower than those inferred from numerical simulations, primarily because the observed sample has significantly higher (0.2-0.4 dex) $L_{\rm p}$ for a given \Mni.
While effects due to composition, mixing, and asymmetry can increase $L_{\rm p}$ current models cannot explain the systematically low $\beta$ values. However, the discrepancy can be alleviated if $\sim$7--50\% of $L_{\rm p}$ for the observed sample originates from sources other than \nickelwospace. Either shock cooling or magnetar spin down could provide the requisite luminosity. Finally, we find that even with our improved measurements, the \Mni\ values of SESN are still a factor of $\sim$3 larger than those of hydrogen-rich Type II SN, indicating that these supernovae are inherently different in terms of their progenitor initial mass distributions or explosion mechanisms

\end{abstract}

\section{Introduction} \label{sec:intro}

Stripped Envelope Supernovae (SESNe) are core-collapse supernovae (SNe) whose progenitors shed a significant fraction of their H envelope before the explosion \citep{Clocchiatti1996,Woosley2002}. It is widely thought that the light curves of SESNe are predominantly powered by the radioactive decay of \nickel synthesized in the explosion \citep{Arnett1982}. In this picture, while shock cooling emission following the often-undetected shock breakout (SBO) may also contribute to the observed luminosity of SESNe during the first few days post-explosion, the main peak of the bolometric light curve is powered by $^{56} \rm{Ni} \rightarrow ^{56}\rm{Co}$ radioactive decay. Following the peak, the light curves of SESNe rapidly decline and subsequently enter a phase of linear (magnitude) decay, which is powered by the $ ^{56}\text{Co} \rightarrow ^{56}\text{Fe}$ chain. This phase typically begins at epochs $\gtrsim$ 60~days \citep{Clocchiatti1997}. The resulting shape of the light curve is not only sensitive to the total mass of \nickelwospace, but also to the total ejecta mass (\Mej), the distribution of \Mni\ within the ejecta, and the degree to which \nickel deposition is asymmetric \citep{Utrobin2017}.

There exists significant diversity within SESNe. Spectroscopically, they are divided into distinct sub-types: IIb, Ib, Ic, and Ic-BL SNe \citep[See][for a review]{Filippenko1997}. The first three sub-types are generally thought to to be produced by increasingly more stripped progenitors \citep{Maund2018}. Type IIb SNe have signatures of both H and He lines, although their H lines are weak and usually disappear after the light curve peak, indicative of a small H mass. Type Ib SNe are SESNe that are H-deficient but exhibit He lines in their spectra, while SNSNe that exhibit neither H nor He lines are categorized as Type Ic SNe. Type Ic-BL SNe are also H- and He-deficient\footnote{Throughout this paper, Type Ic-BL are not included within Type Ic class.}, but are categorized by  broad spectral lines that are indicative of extremely high velocity ejecta  ($\gtrsim 15000$~km~s$^{-1}$; \citealt{Modjaz2014}). They are the only SN sub-type that is associated with long-duration gamma-ray bursts \citep[$l$GRBs;][]{Woosley2006}.

The progenitor systems of SESNe remain a matter of extensive debate. While they are H-poor, their envelopes could, in principle, be removed either via strong stellar winds or via stripping through interaction with a close binary companion \citep{Woosley1995}. In the former case, the progenitors of SESNe would predominately be Wolf-Rayet (WR) stars, with initial masses above 25--30 \Msolar\ \citep[e.g][]{Begelman1986}. In the latter case, many SESNe could be produced by stars with lower initial masses often associated with H-rich Type II SNe (e.g.\ 10$-$20 \Msolar), but that have lost their envelopes via Roche Lobe Overflow (RLOF) prior to explosion (e.g. \citealt{Podsiadlowski1992}).

In recent years, a number of pieces of observational evidence have pointed towards binary stars being a significant contributor to the observed sample of SESNe.
First, binary interaction should be common among stars that are expected to be CC SNe progenitors \citep[e.g.][]{Sana2012} and SESNe constitute about one-third of all core-collapse SNe in volume-limited samples \citep{Li2011,Shivvers2017}. This is higher than the predicted fraction if SESNe solely originate from high-mass WR stars \citep{Smith2011}. Second, unlike H-rich Type II SNe for which dozens of Red Supergiants (RSGs) have been identified in pre-SN images \citep{Smartt2009}, direct progenitor detections of SESNe are scarce \citep{Yoon2012,Eldridge2013}, indicating the progenitors are relatively faint. While a number of Type IIb progenitors have been identified, they are Yellow Supergiants (YSGs). YSGs are not predicted to explode in standard single star evolution models, and thus may indicate close binary progenitor systems \citep{Yoon2017,Sravan2019}. For completely H-stripped SNe, results are even less conclusive. The only reported detections to date are of the progenitors for Type Ic SN2017ein \citep{VanDyk2018,Xiang2019} and Type Ib iPTF13bvn \citep{Cao2013,Kim2015,Eldridge2016}, the former of which has yet to be confirmed. The non-detection of Type Ib/c progenitors in pre-SN images is seemingly in line with the binary scenario where the progenitors are likely to be dim He stars stripped by a companion \citep{Eldridge2013, VanDyk2016}. Lastly, the reported ejecta masses of SESNe are almost exclusively in the range 2--4~\Msolar\ \citep{Drout2011,Lyman2016}. These values are lower than those predicted by models of massive single stars stripped by strong stellar winds ($\gtrsim$6~\Msolar\ for stars with initial masses of 25-150~\Msolar; e.g.\ \citealt{Eldridge2008}), but consistent with expectations for lower initial mass stars stripped in binaries\footnote{Although these results should be interpreted with caution since ejecta masses are often obtained from Arnett-like models, for which some assumptions break down in the case of SESNe as discussed in this paper.}.

However, the conclusion that most SESN are produced by stars from a similar initial mass range as H-rich Type II SNe---simply stripped by a close binary companion---is possibly in tension with other findings. 
The analyses of  H$\alpha$ emission in SN host galaxies reveals that SESNe are more preferably found in star-forming regions compared to H-rich Type II SNe \citep{Anderson2012}. Further studies of stellar populations in the vicinity of SESNe sites indicate that Type IIb, Ib, and Ic SNe are progressively found in younger stellar populations, suggesting that they arise from more massive progenitors \citep{Maund2018}.  In addition, a key piece of evidence that has been particularly problematic for the binary scenario is the reported \nickel masses of SESNe, which are systematically larger than those of H-rich Type II SNe \citep{Anderson2019,Meza2020}. {\nilou This may suggest that the progenitors of SESNe are initially more massive that those of H-rich Type II SNe, which is more naturally predicted by the evolution of single stars}.

Statistical studies of SESNe have reported the average \Mni\ for SESNe $>0.1$~\Msolar\ \citep{Drout2011, Lyman2016, Prentice2016, Prentice2018, Sharon2020}. Recently, \cite{Anderson2019} compiled the \Mni\ of 115 H-rich Type II SNe and 141 SESNe  reported in the literature. They found the average value of \Mni\ for SESNe is 0.293~\Msolar\ which is a factor of $\sim$7 larger than that of H-rich Type II SNe. They argued that this significant discrepancy stems either from differences in the progenitors and explosion mechanisms of Type II versus SESNe or due to systematic errors in the measurement of \Mni\ values from SN light curves that differ between Type II and SESNe.

Indeed, the accuracy of the \Mni\ estimates for SESNe has been disputed in recent years \citep{Dessart2016,Sukhbold2016,Khatami2019, Meza2020}. Unlike H-rich SNe for which \Mni\ is estimated by modeling the radioactive tail of the light curve, the \Mni\ of SESNe is commonly obtained from the peak of their bolometric light curves using the models of \citet{Arnett1980,Arnett1982}. These are a series of widely-used analytical models for radioactive heating/diffusion based on self-similar assumptions that provide bolometric SN light curves and a consequential rule. This ``Arnett's rule'' states that, for SNe powered exclusively by radioactive decay, the radioactive heating rate and the observed bolometric luminosity at the \emph{peak} of the bolometric light curve are equal. Although Arnett's rule roughly holds for Type Ia SNe, the self-similarity condition breaks down when the SN ejecta has a centralized \nickel deposition, calling into question its efficacy when applied to SESNe \citep{Khatami2019}. Thus, alternate means of measuring \Mni\ in SESNe may be required.

{\nilou\ Most directly, \Mni\ for SESNe can be measured by modelling the late-time light curve tail, when the ejecta is optically thin and the luminosity is determined by the instantaneous heating rate. However, these epochs have only been observed for a fraction of known SESN.
\citet{Katz2013} proposed a ``Luminosity Integral'' technique for measuring \Mni\ from radioactively powered SNe, which was recently employed by \cite{Sharon2020} on a dozen SESNe. Although this method does not suffer from many of the simplifying assumptions of Arnett's models, it relies on the temporally well-sampled observations of SN from the explosion epoch to the tail, and will require the addition of an extra parameter if any other power source significantly contributes to the observed luminosity over this timescale.}

Alternatively, \citeauthor{Khatami2019} (\citeyear{Khatami2019}, hereafter, KK19) recently proposed a new analytical model that relates the peak bolometric luminosity and its epoch to the radioactive heating function in order to address the limitations of Arnett's models. However, this model fundamentally depends on the choice of a dimensionless parameter $\beta$ that is sensitive to several physical effects including the spatial distribution of \nickelwospace, the envelope composition, potential explosion asymmetries and extra power sources.
\cite{Meza2020} recently applied the model of KK19 to a sample of SESN, adopting a set of $\beta$ values that were derived from the simulated SESNe light curves of \citet{Dessart2016}.
However, to ascertain if those $\beta$ values are realistic, \emph{they must be directly constrained from observed SESN light curves with independent \Mni\ estimates.} These observationally calibrated $\beta$ values would then offer an independent probe of the progenitors and explosion mechanisms of SESN. In addition, if there exists a robust $\beta$ for each SESN sub-type, then KK19's model can be used, as an alternative to Arnett's rule, {\nilou to give accurate \Mni\ estimates for a large sample of SESNe}. 

In this paper, we present a systematic analysis of nickel mass distribution for 27 well-observed SESNe derived by modeling their radioactive light curve tails, accounting for partial trapping of $\gamma$-rays. The resulting \Mni\ values are then compared against their counterparts measured under Arnett's rule. We also provide a systematic comparison between the nickel masses of SESNe and H-rich Type II SNe, both obtained from the radioactive light curve tail, hence minimizing the biases that originated from the modeling methods in previous studies. In addition, we employ the model of KK19 on the light curves of observed SESNe to 1) calibrate the $\beta$ parameter using the peak light curve properties and our independent tail \Mni\ measurements, and 2) constrain the progenitor and explosion properties of SESNe using our calibrated $\beta$ in comparison to that obtained from numerical simulations.

This paper is organized as follows. In $\S$\ref{sec:nickel}, we describe analytical models that aim to constrain the amount of synthesized \nickel from light curve observables. $\S$\ref{sec:sample} presents our criteria for selecting a sample of well-observed SESNe and a systematic procedure for obtaining their distances, extinction values, bolometric light curves, and explosion epochs.  We provide the \nickel masses and calibrated $\beta$ values for each SN in our sample in \S~\ref{sec:res}. We discuss the implications of our results for understanding their progenitor systems and heating sources in $\S$\ref{sec:disc}, and conclude in $\S$\ref{sec:conc}.

\section{Analytical Models of \Mni} \label{sec:nickel}

To constrain the \Mni\ of SESNe, we employ three analytical models: an optically-thin radioactive decay model for the light curve tail, Arnett's rule for the light curve peak, and KK19's model for the light curve peak. Here, we briefly review their formulation and observational dependencies, as this will influence our SN sample selection in \S~\ref{sec:sample}.
\subsection{Radioactive Decay Modelling of the Light Curve Tail}
The ``light curve tail'' refers to the late-time evolution of the SN light curve once it enters a phase of linear decline in magnitude vs. time. For SESNe, the light curve tail typically begins at an earlier epoch ($t\gtrsim60$~days post-explosion) compared to H-rich Type II SNe, for which the tail is observable only after the H-recombination plateau phase ends ($t\gtrsim90$~days post-explosion). It is widely thought that the tail of core-collapse SNe is powered by the $^{56}\text{Co} \rightarrow ^{56}\text{Fe}$ radioactive decay chain \citep{Colgate1969}. At this stage, the ejecta become transparent to the stored radiative energy; therefore, the observed luminosity traces the instantaneous heating rate. The $\gamma$-rays produced by the radioactive decay heat the ejecta, making the tail of the light curve an appropriate probe for measuring the amount of \nickel produced. The observed luminosity of the tail can be then modeled as: 
\begin{eqnarray}
\label{eq:lni}
&& L \simeq L_\gamma \big(1-e^{-(t/T_0)^{-2}} \big) + L_
 {\rm pos},
\end{eqnarray}
\citep{Wygoda2019}, where $t$ is time since the explosion, $L_\gamma$ is the luminosity produced by the radioactive decay of Co$^{56}$ and Ni$^{56}$, and $L_{\rm pos}$ is the total energy release rate of positron kinetic energy. The term in parenthesis is a deposition factor, which represents the incomplete trapping of $\gamma$-rays with $T_0$ denoting the partial trapping timescale of the tail. The deposition factor is proportional to $1-e^{-t^{-2}}$ for an explosion in homologous expansion \citep{Sutherland1984,Clocchiatti1997}. The luminosity terms in Equation~\ref{eq:lni} can be expressed as:
\begin{eqnarray}
\label{eq:lpos}
&& L_\gamma = M_{\rm Ni} \Big( (\epsilon_{\rm Ni}- \epsilon_{\rm Co})e^{-t/t_{\rm Ni}}+ \epsilon_{\rm Co}e^{-t/t_{\rm Co}}\Big) \\
&& L_{\rm pos} = 0.034 M_{\rm Ni} \epsilon_{\rm Co} \Big( e^{-t/t_{\rm Ni}}- e^{-t/t_{\rm Co}}\Big),
\end{eqnarray}
where $\epsilon_{\rm Ni}= 3.9 \times 10^{10}$~erg~g$^{-1}$~s$^{-1}$ and $\epsilon_{\rm Co}=6.8 \times 10^9$~erg~g$^{-1}$~s$^{-1}$ are the specific heating rates of Ni and Co decay, respectively, and $t_{\rm Ni}=8.8$~days and $t_{\rm Co}=111.3$~days are their corresponding decay timescales. In this formulation, the escape of positrons, which happens on the timescale of several hundred days, is ignored. While the complete trapping of $\gamma$-rays is commonly assumed for H-rich Type II SNe due to their large ejecta masses and correspondingly long $T_0$, it is important to determine the $T_0$ from the slope of light curve tail for SESNe since their ejecta masses are smaller and $T_0$ is usually comparable to the onset time of the radioactive tail. If the bolometric luminosity $L$ can be ascertained observationally, the only unknowns in Equations \ref{eq:lni} and \ref{eq:lpos} are \Mni\ and  $T_0$ which can be determined by fitting the slope and overall normalization of the radioactive tail of the light curve. 

\subsection{Arnett's Rule}
{\nilou While robust, the ``light curve tail'' method is not always accessible because the late-time radioactive tails are often faint, and thus more difficult to observe. As a result, the \Mni\ of SESNe are typically obtained with Arnett's rule,} which states that the instantaneous heating rate from the radioactive decay of \nickel and $^{56}$Co is equal to the bolometric luminosity of the SN at the light curve peak. This can be rewritten as:
\begin{eqnarray}
\label{eq:arnett}
&& L_{\rm p} = M_{\rm Ni} \Big( (\epsilon_{\rm Ni}- \epsilon_{\rm Co})e^{-t_{\rm p}/t_{\rm Ni}}+ \epsilon_{\rm Co}e^{-t_{\rm p}/t_{\rm Co}}\Big), 
\end{eqnarray}
where $t_{\rm p}$ and $L_{\rm p}$ are the peak time and peak bolometric luminosity, respectively.  Arnett-like models make several assumptions to solve the thermodynamic differential equation, including homologous expansion of the ejecta, radiation-dominated pressure, spherical symmetry, and a self-similar energy density profile and also adopt a radiation diffusion approximation \citep{Arnett1980,Arnett1982}.

\subsection{KK19's model}\label{sec:KK19model}
KK19 showed that the assumption of self-similarity for the energy density profile will limit the accuracy of the Arnett-like models, especially for centrally-located heating sources, due to the time-dependent evolution of the diffusion wave through the ejecta. Instead, they propose a new relationship between the peak time, $t_{\rm p}$, and peak luminosity, $L_{\rm p}$, without assuming self-similarity:
\begin{eqnarray}
\label{eq:general}
L_{\rm p}&=& \frac{2}{\beta^2 t_{\rm p}^2} \int_0^{\beta t_{\rm p}} t^\prime L_{\rm heat}(t^\prime) dt^\prime
\end{eqnarray}
where $L_{\rm heat}(t)$ denotes a generic heating function and $\beta$ is a dimensionless parameter of an order of unity. When $L_{\rm heat}(t)$ is powered by  $^{56}\text{Ni} \rightarrow  ^{56}\text{Co} \rightarrow ^{56}\text{Fe}$ radioactive decay, Equation~\ref{eq:general} becomes: 
\begin{eqnarray}
\label{eq:khatami}
L_{\rm p}&=& \frac{2\epsilon_{\rm Ni}M_{\rm Ni} t_{\rm Ni}^2}{\beta^2 t_{\rm p}^2} \Bigg[\Big(1-\frac{\epsilon_{\rm Co}}{\epsilon_{\rm Ni}}\Big)\Big(1-(1+\beta t_{\rm p}/t_{\rm Ni})e^{-\beta t_{\rm p}/t_{\rm Ni}} \Big) \nonumber \\ 
&&+\frac{\epsilon_{\rm Co}t_{\rm Co}^2}{\epsilon_{\rm Ni}t_{\rm Ni}^2}\Big(1-(1+\beta t_p/t_{\rm Co})e^{-\beta t_p/t_{\rm Co}}\Big)\Bigg].
\end{eqnarray}
For this specific form of $L_{\rm heat}(t)$, the following relationships hold: The $M_{\rm Ni}$ required to reproduce a fixed $\{t_{\rm p},L_{\rm p}\}$ pair will be directly proportional to the $\beta$ value adopted.  In contrast, if $M_{\rm Ni}$ is known, then the value of Lp or tp required to reproduce a given $\{t_{\rm p},M_{\rm Ni}\}$ or $\{L_{\rm p},M_{\rm Ni}\}$ pair, respectively, will be \emph{inversely} proportional to the value of $\beta$ adopted (if the light curve is powered entirely by radioactive decay). 

The parameter  $\beta$ incorporates the fact that $L_{p}$ does not necessarily trace the radioactive heating rate as the stored internal energy of ejecta may lag or lead the observed luminosity, $L(t)$, at the time of peak.  The choice of $\beta$ critically depends on several physical effects such as the spatial distribution of  \nickelwospace, the envelope composition, asymmetries in the heating source or ejecta, and all power sources contributing to the observed luminosity along with their exact heating functions. Using Equation~\ref{eq:khatami}, $\beta$ can be derived from the light curve of SESNe with known $L_{\rm p}$ and $t_{\rm p}$, if there is an independent constraint on \Mni. We can then observationally calibrate the appropriate values of $\beta$ using a sample of SESNe of various sub-types. With a sample of calibrated $\beta$ values, one can potentially apply KK19's model to a wide range of SESNe with only photospheric data coverage to constrain their \Mni. In addition, comparing the $\beta$ values obtained from observed SESN light curves to those inferred from numerical light curve models calculated with different input physics can inform us about the explosion details of SESNe.

\section{SN Sample and Methods} \label{sec:sample}

\subsection{Sample Selection} \label{sec:criteria}
Our SESN sample should consist of those SESNe for which a measurement of the \Mni\ can be made from both the light curve tail and the light curve peak. This will allow us to conduct an empirical comparison between the \Mni\ of SESNe obtained from the Arnett model and the radioactive tail and also provides the means to produce a data-driven calibration of the KK19 $\beta$ values for SESN. Therefore, we compiled the photometry of well-observed SESNe in the literature and select SESNe that meet the following criteria:
\begin{enumerate}
    \item well-sampled coverage of the early rise (i.e., multiple observations before 5 days pre-maximum in at list one band) or the observation of an accompanying GRB/X-ray flash, since a constraint on the epoch of explosion is needed to derive \Mni\ from Equations \ref{eq:lni} and \ref{eq:lpos},
    \item multiple photometric measurements on the tail of the light curve, i.e., epochs $\gtrsim$60~days, to be able to obtain \Mni\ from the radioactive tail,
    \item reasonable coverage around the light curve peak, which is required for both computing \Mni\ using the Arnett model (Equation~\ref{eq:arnett}) and calibrating the $\beta$ parameter using KK19's model (Equation~\ref{eq:khatami}),  
    \item light curves in at least two bands over the light curve tail and peak, so that the bolometric luminosity and host galaxy reddening can be computed. (See  \S~\ref{sec:lbol} for the method.) 
\end{enumerate}

We identified 27 SNe from the literature that satisfy the above criteria and downloaded their photometric data from The Open Supernova Catalog \citep{Guillochon2017}. 
Our sample consists of 8 IIb, 8 Ib, 4 Ic, and 7 Ic-BL SNe. These SNe are listed in Table~\ref{tab:sn_basic_param} along with their basic properties, including SN type, the host galaxy name, distance estimate, extinction, and the epoch of explosion.

\begin{deluxetable*}{lllcccc}[ht]
\tabletypesize{\footnotesize}
\tablecolumns{7} 
\tablewidth{2\textwidth}
 \tablecaption{SESN sample with their basic parameters \label{tab:sn_basic_param}}
 \tablehead{
 \colhead{SN name} & \colhead{Host} & \colhead{Type} & \colhead{d (Mpc)} & \colhead{Galactic $E(B-V)$ (mag)} &  \colhead{Host $E(B-V)$ (mag)} & \colhead{t0 (MJD)} 
 } 
\startdata 
 SN1993J & M81 & IIb & 3.6 (0.2)$^a$ & 0.0690 (0.0001) & 0.11 (0.00) & 49074.0 (0.0)$^c$ \\
 SN1994I & M51 & Ic & 8.6 (0.1)$^a$ & 0.0308 (0.0015) & 0.40 (0.04) & 49443.5 (0.3) \\
 SN1996cb & NGC 3510 & IIb & 9.8 (0.7) & 0.0261 (0.0005) & 0.00 (0.01) & 50429.5 (2.0) \\
 SN1998bw & ESO 184-G82 & Ic-BL & 38.1 (2.6) & 0.0494 (0.0011) & 0.00 (0.02) & 50928.9 (0.0)$^d$ \\
 SN2002ap & M74 & Ic-BL & 9.8 (0.5)$^a$ & 0.0616 (0.0018) & 0.01 (0.02) & 52300.0 (2.5) \\
 SN2003jd & MCG -01-59-21 & Ic-BL & 77.9 (5.4) & 0.0378 (0.0005) & 0.00 (0.10) & 52929.0 (2.0) \\
 SN2004aw & NGC 3997 & Ic & 73.6 (5.0) & 0.0184 (0.0009) & 0.37 (0.08) & 53076.5 (6.0) \\
 SN2004gq & NGC 1832 & Ib & 26.3 (1.8) & 0.0629 (0.0004) & 0.12 (0.02) & 53346.1 (4.0) \\
 SN2005hg & UGC 1394 & Ib & 87.7 (6.0) & 0.0894 (0.0021) & 0.52 (0.08) & 53665.6 (2.0) \\
 SN2006T & NGC 3054 & IIb & 34.6 (2.4) & 0.0643 (0.0007) & 0.26 (0.03) & 53764.5 (1.5) \\
 SN2006el & UGC 12188 & IIb & 70.0 (7.0)$^b$ & 0.0975 (0.0012) & 0.07 (0.04) & 53962.3 (0.7) \\
 SN2006ep & NGC 214 & Ib & 61.7 (4.3) & 0.0306 (0.0005) & 0.33 (0.05) & 53970.5 (7.0) \\
 SN2007gr & NGC 1058 & Ic & 9.0 (0.6) & 0.0532 (0.0005) & 0.15 (0.11) & 54329.7 (2.5) \\
 SN2007ru & UGC 1238 & Ic-BL & 60.9 (4.2) & 0.2217 (0.0046) & 0.00 (0.02) & 54429.5 (3.0) \\
 SN2007uy & 	NGC 2770 & Ib & 31.4 (2.2) & 0.0192 (0.0003) & 0.53 (0.02) & 54464.0 (3.5) \\
 SN2008D & NGC 2770 & Ib & 31.4 (2.2) & 0.0193 (0.0002) & 0.47 (0.04) & 54474.6 (0.0)$^e$ \\
 SN2008ax & NGC 4490 & IIb & 9.2 (0.6) & 0.0188 (0.0002) & 0.25 (0.04) & 54528.3 (1.0) \\
 SN2009bb & NGC 3278 & Ic-BL & 40.1 (2.8) & 0.0847 (0.0010) & 0.40 (0.08) & 54912.9 (1.1) \\
 SN2009jf & NGC 7479 & Ib & 33.8 (2.4) & 0.0970 (0.0013) & 0.07 (0.06) & 55099.5 (4.2) \\
 SN2011bm & IC 3918 & Ic & 99.2 (6.8) & 0.0285 (0.0005) & 0.00 (0.15) & 55645.5 (0.5) \\
 SN2011dh & M51 & IIb & 8.6 (0.1)$^a$ & 0.0309 (0.0017) & 0.15 (0.03) & 55712.0 (0.0)$^c$ \\
 iPTF13bvn & NGC 5806 & Ib & 23.9 (1.7) & 0.0436 (0.0006) & 0.15 (0.04) & 56458.3 (0.8) \\
 SN2013df & NGC 4414 & IIb & 17.9 (1.0)$^a$ & 0.0168 (0.0002) & 0.20 (0.04) & 56447.3 (0.9)$^c$ \\
 SN2013ge & NGC 3287 & Ib & 23.7 (1.6) & 0.0198 (0.0002) & 0.10 (0.10) & 56602.3 (4.7) \\
 SN2014ad & PGC 37625 & Ic-BL & 26.7 (1.9) & 0.0380 (0.0012) & 0.10 (0.07) & 56724.5 (3.0) \\
 SN2016coi & UGC 11868 & Ic-BL & 18.1 (1.3) & 0.0737 (0.0021) & 0.27 (0.08) & 57533.2 (2.1) \\
 SN2016gkg & NGC 613 & IIb & 19.7 (1.4) & 0.0166 (0.0002) & 0.20 (0.17) & 57655.2 (0.0)$^c$ \\
\enddata
\tablenotetext{a}{Redshift-independent distances. SN1993J and 2013df \citep{Gerke2011}; SN1994I and 2011df \citep{McQuinn2016}; SN2002ap \citep{McQuinn2017}}
\vspace{-0.1in}
\tablenotetext{b}{Distance from the value reported in \cite{Drout2011} }
\vspace{-0.1in}
\tablenotetext{c}{Epoch of explosion from the shock cooling emission}
\vspace{-0.1in}
\tablenotetext{d}{Epoch of explosion from the GRB emission}
\vspace{-0.1in}
\tablenotetext{e}{Epoch of explosion from the X-ray emission}
\vspace{-0.05in}
 \tablecomments{References: SN1993J \citep{Richmond1994,Richmond1996}; SN1994I \citep{Richmond1996_94I}; SN1996cb \citep{Qiu1999}; SN1998bw \citep{Galama1998,McKenzie1999}; SN2002ap \citep{Pandey2003,Yoshii2003}; SN2003jd \citep{Valenti2008}; SN2004aw \citep{Taubenberger2006}; SN2004gq \citep{Bianco2014, Stritzinger2018b}; SN2005hg \citep{Drout2011}; SN2006T \citep{Stritzinger2018b}; SN2006el \citep{Drout2011,Bianco2014}; SN2006ep \citep{Bianco2014,Stritzinger2018b}; SN2007gr \citep{Hunter2007}; SN2007ru \citep{Sahu2009}; SN2007uy \citep{Bianco2014}; SN2008D \citep{Bianco2014}; SN2008ax \citep{Pastorello2008}; SN2009bb \citep{Pignata2011}; SN2009jf \cite{Sahu2011}; SN2011bm \citep{Valenti2012}; SN2011dh \citep{Tsvetkov2012}; iPTF13bv \citep{Folatelli2016,Fremling2016}; SN2013df \citep{Morales2014,Shivvers2019}; SN2013ge \citep{Drout2011}; SN2014ad \citep{Sahu2018}; SN2016coi \citep{Prentice2018}; SN2016gkg \citep{Bersten2018};
 }
\end{deluxetable*}

\subsection{Distances} \label{sec:distance}
The distances we adopt throughout our analyses are listed in Table~\ref{tab:sn_basic_param}. We adopt up-to-date host galaxy distances reported in NASA/IPAC Extragalactic Database (NED)\footnote{\url{https://ned.ipac.caltech.edu/}}. We prioritize distances obtained by Cepheids and Tip of Red Giant Branch methods when available and otherwise use cosmology-dependent values. For redshift-dependent distances, we adopt the standard $\rm{\Lambda CDM}$ cosmology with a Hubble constant $H_0=73.24$~km~s$^{-1}$~Mpc$^{-1}$, matter density parameter $\Omega_M=0.27$, vacuum  density parameter $\Omega_\Lambda=0.73$ \citep{Riess2016} and correct for Virgo, Great Attractor, and Shapley Supercluster Infall. For one object (SN2006el, which exploded in the galaxy UGC 12188), no host galaxy redshift is listed in NED. We therefore adopt the distance given in \cite{Drout2011}, which is based on a host galaxy redshift reported in ATel 854 \citep{Antilogus2006}. {We note that adopting Planck cosmological parameters \citep[i.e., $H_0=67.4$~km~s$^{-1}$~Mpc$^{-1}$, $\Omega_M=0.315$, and $\Omega_\Lambda=0.685$,][]{PlanckCollaboration2018} would increase redshift-dependent distances ($\sim$80\% of our sample) by $\sim 8$\%. Possible systematic effects of this choice on our results will be discussed in \S~\ref{sec:disc} below.}

\subsection{Galactic and Host Galaxy Extinction} \label{sec:extinction}
We adopt the value for galactic extinction along the line of sight to each SN reported in  NASA/IPAC Infrared Science Archive\footnote{\url{https://irsa.ipac.caltech.edu/applications/DUST/}} based on the extinction model of \cite{Schlafly2011} and assuming a $R_V=3.1$ extinction law. The resulting values are listed in Table~\ref{tab:sn_basic_param}.

To estimate values for host galaxy extinction, we use the intrinsic SN color-curve templates of \cite{Stritzinger2018} and attribute the difference between the observed and the intrinsic color to the host galaxy reddening. Specifically, the host extinction can be written in terms of the observed minus intrinsic color as: $ E(X-Y)_{\rm{host}}=(X-Y)_{\rm obs}-(X-Y)_{\rm int}$, where $X$ and $Y$ are the measured magnitudes corrected for the Galactic extinction in two different filters. When computing the extinction, we take the average of the color difference between the observed data and the templates of \cite{Stritzinger2018} from 5 days to 10 days post-maximum. When determining this average, time of maximum is defined based on the observed filter that we adopt as the ``$X$''-band in the above expression. 

Whenever available, we use $X-Y=V-R/r/i$ color indices. Since there is no template provided for $V-R$ intrinsic color in \cite{Stritzinger2018}, we convert observed Johnson $R$-band photometry to Sloan $r$-band using the color transformation relation of \cite{Jordi2006} when required. $E(X-Y)$ is then converted to the standard reddening $E(B-V)$ using the bandpass coefficients of \cite{Schlafly2011} assuming an $R_V=3.1$ Milky Way extinction law. For those SNe for which photometric data is not available in any of the $R/r/i$ filters, we adopt $B-V$ color index instead. Furthermore, we find that our obtained reddening is not robust to the choice of filters $X$ and $Y$ for several Type IIb SNe (i.e., SN1993J, SN2011dh, and SN2013df). For these SNe, we take the average of $E(B-V)$ values derived using different color indices such as $V-r$, $V-i$, and $B-V$. The final host galaxy extinctions used in our analyses are listed in Table~\ref{tab:sn_basic_param}. 

{It worth noting that \cite{Stritzinger2018} constrained $R_V$ directly for a set of 8 SESN with both optical and IR light curves and moderate reddening. They found values spanning a wide range (1.1 $\lesssim$ $R_V$ $\lesssim$ 4.3), with tentative evidence for Type Ic SN showing higher $R_V $values than Type IIb/Ib SN. We continue to adopt a standard $R_V=3.1$ Milky Way
extinction law  here, due primarily to the small number of SESN for which this has been directly constrained, as well as for consistency with previous literature extinction estimates. However, we note that extinction corrections in the $B/V/r/i$ bands considered here could vary by $\sim$0.05--0.3 mag 
for $R_V$ values between 1.1 and 4.3. Possible systematic effects of this choice on our results will be discussed in \S~\ref{sec:disc} below.}


\subsection{Epoch of Explosion} \label{suc:texp}
The estimated epochs of explosion for each SN are presented in Table~\ref{tab:sn_basic_param}. For several of Type IIb SNe with double-peaked light curves, the epoch of explosion is adopted from the reported values obtained from by modeling the shock cooling emission. For SN1998bw, which is a Type Ic-BL SN associated with GRB 980425, we take the GRB epoch as the explosion epoch. Similarly, for SN2008D, the epoch of the observed X-ray flash XRO080109 is taken as the explosion epoch. {\nilou Since the shock velocities of these SNe are high, we can assume that the GRB or X-ray flash occurs shortly after the explosion time; therefore, GRB 980425 and XRO080109 provide accurate estimates of the explosion epochs.} For the rest of SNe in our sample, we estimate the explosion time by fitting a power-law with the form:
\begin{eqnarray}
f &=& \begin{cases}
f_0 (t-t_0)^n  &\hspace{1cm} t>t_0\\
0 &\hspace{1cm}  t\leq t_0
\end{cases},
\end{eqnarray}
to the observed fluxes, $f$, in the band with the best early light curve coverage. We carry out least-square regression to fit for the power index $n$, scaling coefficient $f_0$, and epoch of explosion $t_0$. For the fitting, we only consider epochs that are pre-maximum-light and within 5 days of the first reported detection as well as any publicly available non-detection upper limits. The uncertainties on the explosion epoch quoted in Table~\ref{tab:sn_basic_param} come directly from the fitting process. The consequences of a possible ``dark phase'' \citep{Piro2013} between the explosion epoch and the epoch of first light will be discussed in \S~\ref{sec:dark}.

\begin{deluxetable*}{lcccccccccc}[ht]
\tabletypesize{\footnotesize}
\tablecolumns{9} 
\tablewidth{2\textwidth}
 \tablecaption{The fit parameters for the SESN sample \label{tab:sn_derived_param}}
 \tablehead{
 \colhead{SN name} &\colhead{Type} & \colhead{BC bands} & \colhead{$\log \ L_{\rm p}$ (erg ~s$^{-1}$)} & \colhead{$ t_{\rm p}$ (days)} & \colhead{Tail \Mni\ (\Msolar)} &  \colhead{$ T_0$ (days)} & \colhead{Arnett  \Mni\ (\Msolar) } & \colhead{Calibrated $\beta$} & \colhead{KK19  \Mni\ (\Msolar)$^{1}$ }& \colhead{$f$ }
 } 
\startdata 
SN1993J & IIb &$ V-I$&42.41 (41.56)&22.0 (0.0)&0.081 (0.005)&110.9 (9.9)&0.15 (0.02)&0.80 (0.16)&0.80 (0.16)&0.23\\
SN1994I&Ic&$ V-I$&42.62 (41.56)&8.2 (0.3)&0.048 (0.015)&57.0 (10.3)&0.11 (0.01)&0.00 (0.00)&0.080 (0.007) &0.48\\
SN1996cb&IIb&$ V-R$&41.87 (41.12)&21.7 (2.0)&0.030 (0.003)&96.8 (14.6)&0.04 (0.01)&1.17 (0.27)&0.023 (0.004) &-0.01\\
SN1998bw&Ic-BL&$ V-I$&43.16 (42.38)&16.6 (0.0)&0.300 (0.013)&108.9 (14.5)&0.67 (0.11)&0.50 (0.19)&0.312 (0.052) &0.38\\
SN2002ap&Ic-BL&$ V-I$&42.47 (41.60)&13.1 (2.5)&0.062 (0.003)&113.4 (12.3)&0.11 (0.02)&0.64 (0.20)&0.058 (0.008) &0.27\\
SN2003jd&Ic-BL&$ V-R$&42.85 (42.11)&15.0 (2.1)&0.117 (0.014)&101.4 (16.1)&0.29 (0.06)&0.29 (0.21)&0.145 (0.026) &0.47\\
SN2004aw&Ic&$ V-I$&42.71 (41.93)&15.6 (6.0)&0.151 (0.030)&150.1 (92.0)&0.22 (0.07)&1.03 (0.26)&0.133 (0.022) &0.08\\
SN2004gq&Ib&$ B-V$&42.36 (41.67)&12.3 (4.0)&0.044 (0.008)&144.6 (74.2)&0.08 (0.03)&0.59 (0.32)&0.051 (0.011) &0.29\\
SN2005hg&Ib&$ V-R$&43.09 (42.34)&18.8 (2.0)&0.353 (0.039)&143.2 (19.2)&0.63 (0.13)&0.83 (0.23)&0.348 (0.063) &0.21\\
SN2006T&IIb&$ B-V$&42.63 (42.00)&15.5 (1.5)&0.082 (0.012)&143.6 (29.5)&0.18 (0.05)&0.49 (0.30)&0.105 (0.025) &0.39\\
SN2006el&IIb&$ V-r^2$&42.27 (41.63)&21.4 (0.7)&0.052 (0.005)&126.4 (21.2)&0.11 (0.02)&0.70 (0.25)&0.057 (0.013) &0.31\\
SN2006ep&Ib&$ B-V$&42.55 (41.83)&15.0 (7.0)&0.058 (0.015)&112.2 (31.1)&0.16 (0.07)&0.29 (0.22)&0.088 (0.017) &0.47\\
SN2007gr&Ic&$ V-I$&42.38 (41.60)&9.8 (2.5)&0.047 (0.004)&126.9 (15.7)&0.07 (0.02)&0.77 (0.33)&0.049 (0.008) &0.18\\
SN2007ru&Ic-BL&$ V-I$&42.90 (42.13)&11.1 (3.0)&0.103 (0.009)&107.7 (14.9)&0.27 (0.07)&0.09 (0.13)&0.147 (0.025) &0.50\\
SN2007uy&Ib&$ B-V$&42.80 (42.16)&16.1 (3.5)&0.143 (0.024)&119.9 (25.4)&0.29 (0.09)&0.64 (0.29)&0.166 (0.037) &0.31\\
SN2008D&Ib&$ V-r^2$&42.08 (41.32)&19.9 (0.2)&0.036 (0.003)&104.6 (14.1)&0.07 (0.01)&0.82 (0.21)&0.036 (0.006) &0.22\\
SN2008ax&IIb&$ V-R$&42.31 (41.58)&21.8 (1.2)&0.060 (0.009)&99.2 (17.2)&0.12 (0.02)&0.73 (0.21)&0.063 (0.012) &0.29\\
SN2009bb&Ic-BL&$ V-I$&42.78 (42.01)&11.0 (1.1)&0.158 (0.055)&47.8 (14.5)&0.19 (0.04)&1.23 (0.35)&0.109 (0.019) &-0.04\\
SN2009jf&Ib&$ V-I$&42.68 (41.91)&22.3 (4.2)&0.164 (0.022)&158.3 (22.7)&0.29 (0.07)&0.87 (0.20)&0.155 (0.026) &0.19\\
SN2011bm&Ic&$ V-I$&42.86 (42.09)&34.6 (0.7)&0.574 (0.038)&119.3 (90.3)&0.63 (0.11)&1.71 (0.41)&0.336 (0.057) &-0.33\\
SN2011dh&IIb&$ V-R$&42.49 (41.48)&20.1 (0.0)&0.093 (0.002)&102.0 (8.6)&0.17 (0.02)&0.81 (0.12)&0.090 (0.009) &0.22\\
iPTF13bvn&Ib&$ V-I$&42.31 (41.54)&16.7 (0.8)&0.040 (0.006)&134.9 (15.6)&0.10 (0.02)&0.43 (0.19)&0.055 (0.009) &0.42\\
SN2013df&IIb&$ V-I$&42.37 (41.53)&22.6 (1.6)&0.063 (0.004)&128.2 (12.6)&0.14 (0.02)&0.61 (0.14)&0.074 (0.011) &0.37\\
SN2013ge&Ib&$ B-V$&42.39 (41.78)&16.8 (4.7)&0.063 (0.012)&122.8 (31.3)&0.11 (0.04)&0.79 (0.33)&0.065 (0.016) &0.24\\
SN2014ad&Ic-BL&$ V-I$&42.84 (42.07)&16.4 (3.0)&0.147 (0.023)&96.7 (15.6)&0.32 (0.08)&0.54 (0.20)&0.148 (0.025) &0.36\\
SN2016coi&Ic-BL&$ V-I$&42.84 (42.06)&17.6 (2.1)&0.170 (0.016)&135.7 (18.3)&0.34 (0.07)&0.67 (0.20)&0.153 (0.026) &0.30\\
SN2016gkg&IIb&$ V-I$&42.29 (41.51)&16.5 (4.0)&0.055 (0.004)&124.8 (15.5)&0.09 (0.02)&0.91 (0.24)&0.050 (0.008) &0.15\\
\enddata
\tablenotetext{1}{\Mni\ values obtained from KK19's model assuming the median values of calibrated $\beta$ reported in Table~\ref{tab:beta_stat} (see \S~\ref{sub:khatami}). }
\vspace{-0.1in}
\tablenotetext{2}{Since the BCs provided by \cite{Lyman2014} are given in Johnson bands, magnitudes in Sloan $r$ are first converted to Johnson $R$ using the transformations of \cite{Jordi2006}. }
\end{deluxetable*}

\begin{figure*}[t!]
\centering
\includegraphics[trim=0.0cm 0.2cm  0cm 0.2cm, clip=true, scale=0.56]{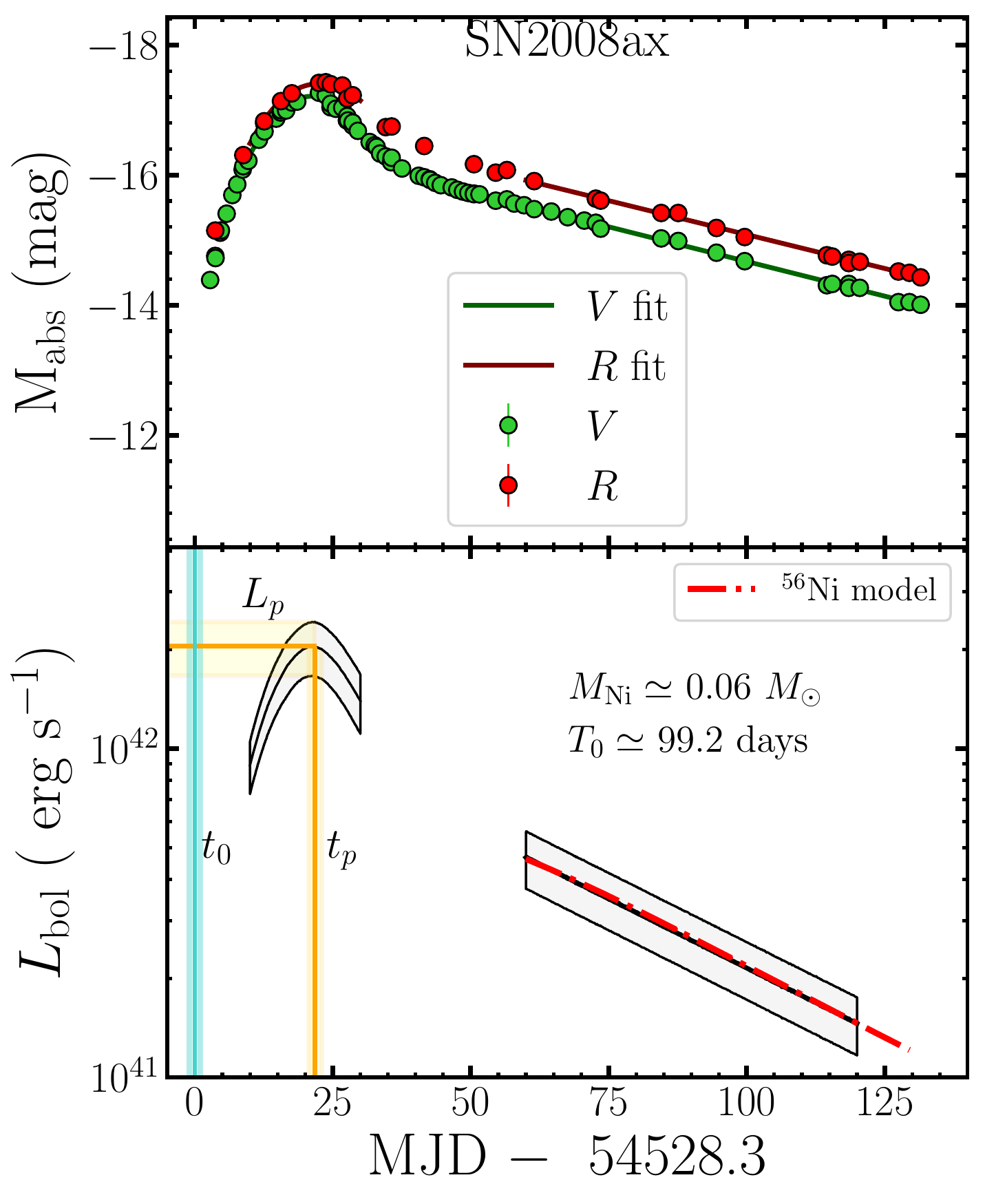}
\includegraphics[trim=0.0cm 0.2cm  0cm 0.2cm, clip=true, scale=0.56]{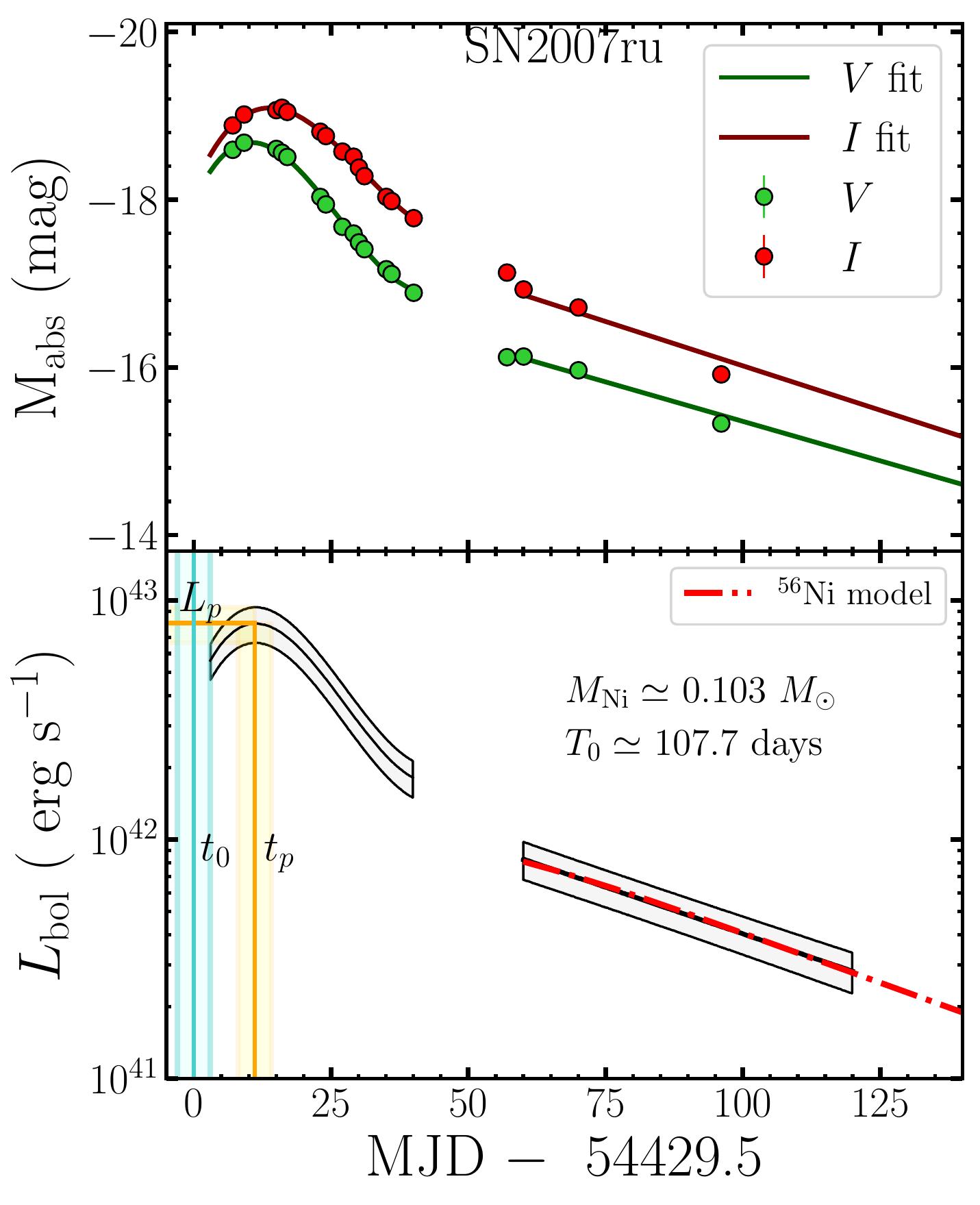}
\caption{Extinction-corrected absolute magnitude (top panel) and bolometric luminosity (bottom panel) light curves for SN2008ax (left panel) and SN2007ru (right panel). The green and red solid curves represent fits to the absolute magnitudes in $V$ and $R/I$ bands, respectively. The cyan vertical line indicates the epoch of explosion $t_0$, while the vertical and horizontal orange lines denote the peak time $t_{\rm p}$ and peak luminosity $L_{\rm p}$ of the bolometric light curves, respectively.  The 1-$\sigma$ confidence level in $t_0$, $t_{\rm p}$, and $L_{\rm p}$ are shown with cyan and orange strips. The dot-dashed red curve in the lower panels represents the best-fit \nickel model of Equation~\ref{eq:lni} to the bolometric radioactive tails (see annotation for fit parameters).  Note when error bars are not visible in the top panel they are smaller than the plotted points. }
\label{fig:snexam}
\end{figure*}

\subsection{Bolometric Light Curves} \label{sec:lbol}
The relative paucity of SESN with extensive coverage in UVOIR bands from early to late times is a challenge for computing bolometric light curves that are needed to obtain \Mni. Here, we focus on obtaining the bolometric luminosities for epochs around the light curve peak and the late-time tail. In order to leverage the multi-band photometric data available for our SESNe sample, we adopt the bolometric correction (BC) coefficients of \cite{Lyman2014, Lyman2016}. These color-dependent coefficients were measured by fitting the SEDs of a sample of SESNe that have coverage in ultra-violet, optical, and infrared wavelengths, and can be utilized as long as light curves for a SN of interest are available in a minimum of two bands. 

We first compute the absolute magnitude light curves for all SN in our sample in the pair of bands indicated in Table~\ref{tab:sn_derived_param} (column ``BC bands''). We correct for the distances and total line-of-sight extinction described above. Next, we fit the multi-band absolute magnitude light curves around the peak and over the radioactive tail with a spline-smoothing function and linear function, respectively. We perform a Monte Carlo (MC) analysis to propagate the uncertainties in the measured magnitudes and distance estimate. The fitted absolute magnitude light curves, which together also provide intrinsic color as function of time, are then used to calculate a bolometric magnitude light curve by applying the color-dependent BC polynomials of \cite{Lyman2014, Lyman2016}. Specifically, ${\rm M}_{\rm bol} = {\rm M}_X + {\rm BC}$, where the  BC is computed using the color indices listed in column ``BC bands'' of Table~\ref{tab:sn_derived_param}, and $X$ denotes the first indicated band listed in that column. Finally, we convert bolometric magnitudes to luminosities assuming M$_{\text{bol},\Sun}=4.74$ mag and $L_{\text{bol},\Sun}=3.83 \times 10^{33}$ erg s$^{-1}$.

Figure~\ref{fig:snexam} illustrates this process. It presents the absolute magnitude (top panel) and the resulting bolometric  (bottom panel) light curves for two SNe in our sample: SN2008ax and SN2007ru. These two objects were specifically chosen to span the range of light curve coverage available during the early rise and late-time tail for SN in our sample.
The spline and linear fits to the absolute magnitude light curves are shown in the top panel. The x-axis represents time since the inferred epoch of explosion derived in \S~\ref{suc:texp}. Gray regions indicate the uncertainties in the bolometric luminosity obtained at each epoch. These uncertainties stem from (in decreasing order of importance): error in the distance estimate, BC  error, and photometric error. We also mark the epoch of explosion $t_0$, the peak luminosity $L_{\rm p}$ and peak time $t_{\rm p}$ of the bolometric light curve in the bottom panel, with shaded regions representing the uncertainty on each parameter. 

\begin{figure}[ht]
\centering
\includegraphics[trim=0.4cm 0.0cm  0.5cm 0cm, clip=true, scale=0.6]{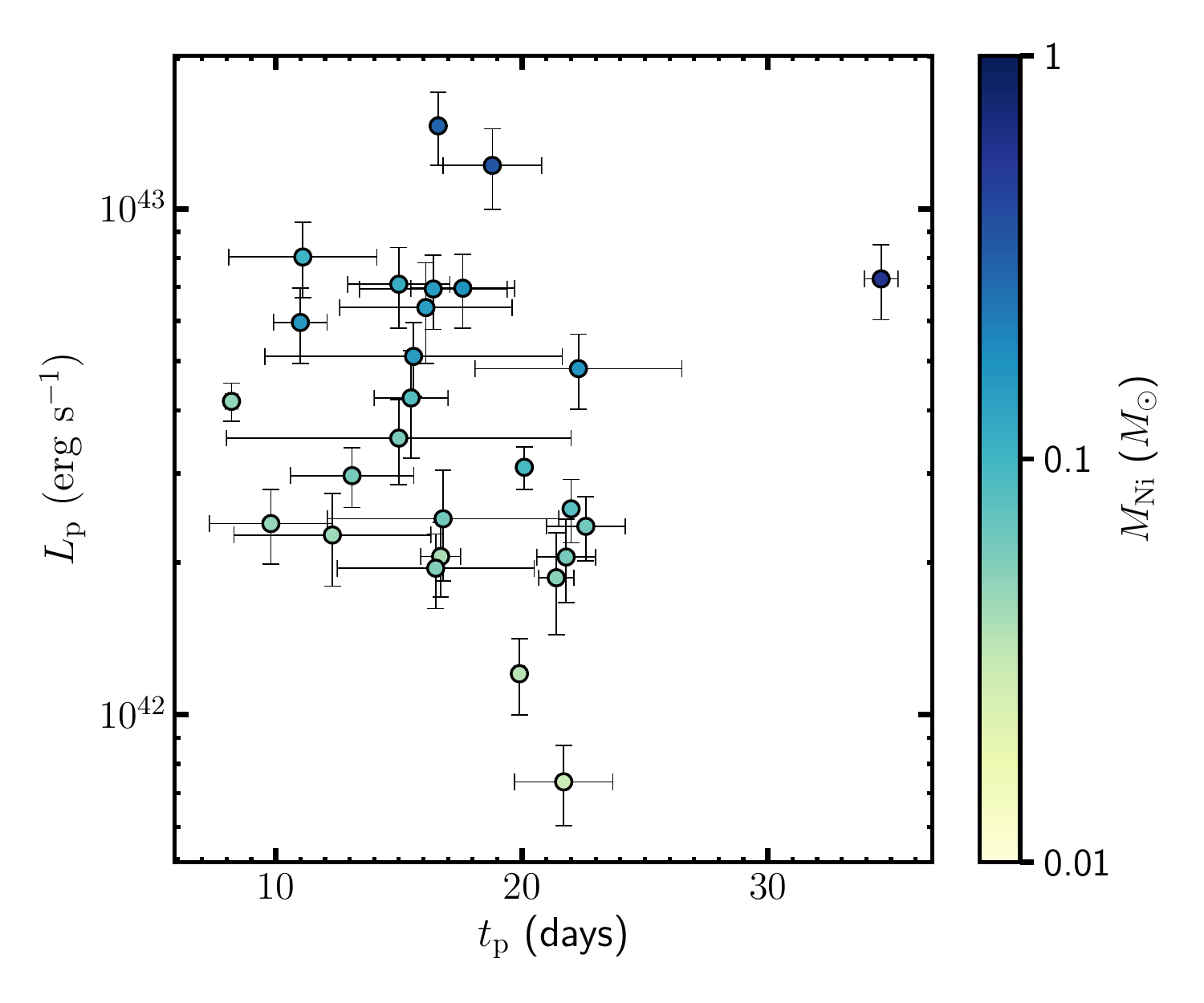}
\caption{Peak time, $t_{\rm p}$, versus peak luminosity, $L_{\rm p}$, for our sample of SESNe. Markers are color-coded based on their tail \Mni\ value.}
\label{fig:tp_vs_lp}
\end{figure}

Derived values of $L_{\rm p}$ and $t_{\rm p}$ for each SN are listed in Table~\ref{tab:sn_derived_param}, and plotted in Figure~\ref{fig:tp_vs_lp}. Note that the final error in $t_{\rm p}$ is a combination of the error in the $t_0$ and bolometric light curve. Overall, the peak luminosities for our sample span $10^{41.87}$--$10^{43.09}$~erg~s$^{-1}$ and rise times span 8.2--34.6 days (SN1994I and SN2011bm, respectively). As seen in Figure~\ref{fig:tp_vs_lp}, no correlation is  apparent between $L_{\rm p}$ and $t_{\rm p}$ for the SESNe in our sample. 

Comparing our $L_{\rm p}$ estimates with previous studies, we find that  our derived peak luminosities are a factor $\sim$2 larger than those found by \cite{Meza2020} who integrate luminosity over the BVRIJH bands, but they are in good agreement with those of \cite{Prentice2016} who find peak luminosities by summing over the UBVRIJHK bands for a fraction of their SN sample. In addition, our $L_{\rm p}$ estimates are consistent within our quoted errors with those of \cite{Lyman2016}, who also adopt the BC polynomial fits of \cite{Lyman2014}.

\section{Results} \label{sec:res}

\subsection{Nickel Masses from the Radioactive Tail}\label{sub:tail}
To constrain \Mni\ of our SESNe sample, we model their radioactive light curve tails using the analytic model of \cite{Wygoda2019} discussed in \S~\ref{sec:nickel}. This model is similar to those of \cite{Valenti2008} and \cite{Drout2013} with one minor modification: the positrons' escape is neglected. Since positrons' escape occurs on a time scale of a few thousand days, it should not affect our results.

By fitting the bolometric luminosity and slope of the radioactive tail for the SESN sample, we can constrain the two unknown parameters in Equation~\ref{eq:lni}: the nickel mass, \Mni, and partial trapping timescale of the tail, $T_0$. The fitting is done in an MC fashion: we run 1000 trials drawing from the distribution of possible luminosities and epochs of explosion. This allows us to propagate the uncertainties in these quantities when obtaining \Mni\ and $T_0$. We only consider epochs of $\geq 60$~days post-explosion, when the ejecta of SESNe are expected to be optically thin, such that the bolometric luminosity will be set by the instantaneous heating rate. In Figure~\ref{fig:snexam}, we display the best-fit radioactive tail models for SN\,2008ax and SN\,2007ru (dot-dashed red curves; bottom panels). As shown, the model closely matches the evolution of the bolometric radioactive tail.
The best-fit parameters, ``tail \Mni'' and $T_0$, for each SN are listed in Table~\ref{tab:sn_derived_param}. Our best-fit tail \Mni\ values range from $\sim$0.030~\Msolar\ to $\sim$0.574~\Msolar\ with a median value of 0.08~\Msolar. $T_0$ is in the range $\sim$47.8--158.3~days with a median value of 116.6~days. The points in Figure~\ref{fig:tp_vs_lp} are color-coded based on these derived tail \Mni\ values. Objects with larger \Mni\ values also exhibit brighter peak luminosities, as expected for SNe powered predominately by radioactive decay.

\begin{figure}[t!]
\centering
\includegraphics[trim=0.2cm 0.2cm  0.2cm 0.2cm, clip=true, scale=0.6]{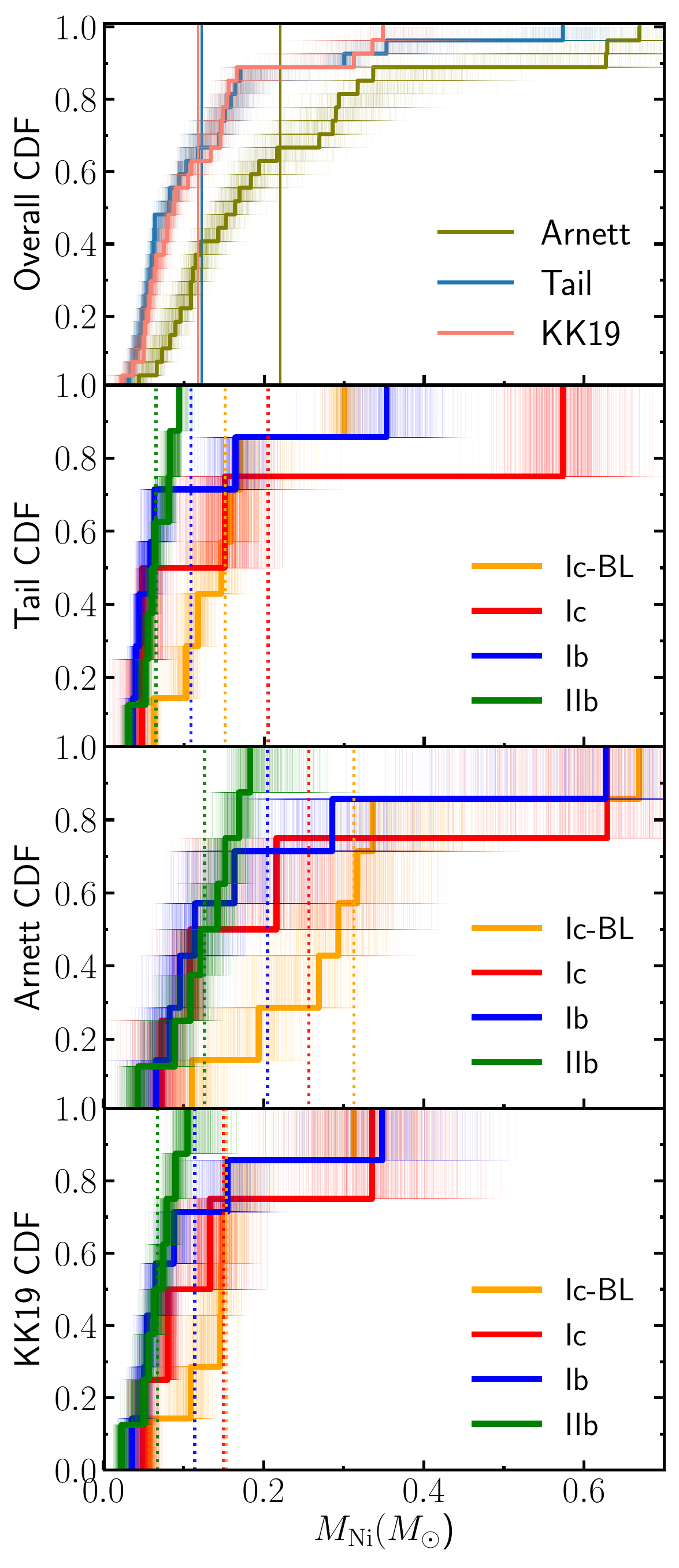}
\caption{Cumulative Distribution Functions of \Mni. \emph{Top:} The \Mni\ CDFs obtained using the radioactive tail modeling (blue curve), Arnett's rule (olive curve), and KK19 model (pink curve). \emph{Middle top:} The tail \Mni\ CDFs categorized into the SN types: Type Ic-BL (yellow curve), Ic (red curve), Ib (blue curve), and IIb (green curve). \emph{Middle bottom \& bottom:} same as the middle top panel but for Arnett and KK19 CDFs, respectively. KK19 \Mni\ values are obtained using the median values of calibrated $\beta$ (see \S~\ref{sub:khatami}). The vertical lines show the mean of each distribution. {\nil The hatched regions represent the uncertainties in CDFs.}}
\label{fig:cdf}
\end{figure}

\begin{deluxetable*}{l|ccc|ccc|ccc}[t!]
\label{tab:mni_stat}
\tabletypesize{\footnotesize}
\tablecolumns{10}
\tablewidth{0pc}
\tablecaption{Table of \nickel Mass Statistics }
\tablehead{\colhead{}& \multicolumn{3}{|c|}{Tail \Mni\ (\Msolar)} &\multicolumn{3}{c|}{Arnett \Mni\ (\Msolar)}& \multicolumn{3}{c}{KK19 \Mni\ (\Msolar)}   \\
{SN Type} & {Mean} & {Median} & {Std} &  {Mean} & {Median} & {Std} &  {Mean} & {Median} & {Std}
}
\startdata
IIb &       0.06 &  0.06 &  0.02 &  0.13 &  0.13 &  0.04 & 0.07 & 0.07 & 0.02\\
Ib &       0.11 &  0.06 &  0.11 &  0.20 &  0.11 &  0.19 & 0.12  & 0.08  & 0.1
\\
Ic &       0.20 &  0.10 &  0.22 &  0.26 &  0.16 &  0.22 
& 0.15  & 0.11 & 0.11\\
Ic-BL &       0.15 &  0.15 &  0.07 &  0.31 &  0.29 &  0.16 & 0.15 & 0.15 & 0.07\\
All &       0.12 &  0.08 &  0.12 &  0.22 &  0.16 &  0.17 & 0.12  & 0.09 & 0.09
 \\
\enddata
\end{deluxetable*} 

We report the basic statistics of our results (mean, median, and standard deviation) for both the full sample and separated by SESN sub-type in Table~\ref{tab:mni_stat}. {\nil However, we note that our sample size is relatively small when SNe are categorized by sub-types; especially normal Type Ic SNe, for which only 4 events met all of our sample criteria outlined in \S~\ref{sec:criteria} and whose distribution may be skewed by the extreme event SN\,2011bm. Therefore, we conduct Analysis of Variance (ANOVA) test to check whether the reported differences between the mean \Mni\ of SN sub-types are statistically significant. The result of ANOVA test indicates that the pairwise comparison of \Mni\ between SN sub-types is not generally statistically significant. One exception for Type Ic-BL SNe for which the reported mean \Mni\ was found to be higher than that of the combined sample of all other SESN sub-types with p-value <0.05.  This is consistent with previous studies which have typically found systematically higher \Mni\ for Type Ic-BL events \citep[e.g.][]{Drout2011,Lyman2016,Prentice2016}. }

Figure~\ref{fig:cdf} displays Cumulative Distribution Functions (CDFs) of the derived \Mni\ for our sample of SESNe. In the top panel of Figure~\ref{fig:cdf}, the \Mni\ CDFs are provided for the entire SESN sample obtained using multiple methods: radioactive tail modeling (blue), Arnett's rule (green), and KK19 model (pink). (See \ref{sub:comp} and \ref{sub:mnikk19} for Arnett and KK19 methods, respectively.) {\nil In order to account for the errors in individual \Mni\ measurements when plotting the CDFs, we run 1000 MC trials in which we sample each \Mni\ value based on the distribution defined by its uncertainly and construct a new CDF. These sampled CDFs are over-plotted in Figure~\ref{fig:cdf}, forming hatched regions that represent the uncertainties associated with the obtained CDFs.}

In the second panel of Figure~\ref{fig:cdf} we present the CDFs of the tail \Mni\ values for each SESN sub-types separately. We conduct Kolmogorov-Smirnov (K-S) tests on the CDFs of tail \Mni\ estimates for each sub-type. A K-S test on the CDF of Type Ic-BL and Ib/c SNe rejects the null hypothesis that these SN types are drawn from the same groups of explosions with a p-value~$=0.02$. In contrast, we find that Type Ic and Ib SNe are likely to be drawn from the same distributions with p-value~$=0.24$. Similarly, a K-S test on Type IIb and Ib/c SNe supports the null hypothesis that these samples originate from the same distribution with p-value~$=0.10$. The p-value further increases to 0.45 when we compare the CDF of Type IIb and Ib SNe.

Radioactive tail nickel masses have previously been estimated for a number of SN in our sample, and results are consistent. In particular, the tail \Mni\ values reported in the recent work of  \cite{Meza2020} are lower limits as they do not take the partial trapping of $\gamma$-rays into account. Our tail \Mni\ estimates provided in  Table~\ref{tab:sn_derived_param} are consistent with these lower limits for SNe that are shared between both samples. {\nilou Similarly, our \Mni\ and $T_0$ estimates are consistent with (within the margin of error) those obtained from the Katz Integral method \citep{Sharon2020} for 8 SESNe in common between the samples.}

\subsection{Comparison to Arnett Model}
\label{sub:comp}
For comparison, we also measure \Mni\ using Arnett's rule, described in \S~\ref{sec:nickel}, which has been extensively employed in the literature. For each SN in our sample,  we fit for \Mni\ in Equation~\ref{eq:arnett} assuming the $t_{\rm p}$ and $L_{\rm p}$ values listed in Table~\ref{tab:sn_derived_param}. Similar to the procedure of deriving tail \Mni, we run 1000 MC trials to take into account the uncertainties in $t_{\rm p}$ and $L_{\rm p}$ when obtaining Arnett \Mni\ values. Results for individual SN are provided in Table~\ref{tab:sn_derived_param}, while basic statistics of both the full Arnett distribution and SESN sub-types are reported in \ref{tab:mni_stat}. The Arnett \Mni\ values span  0.04~\Msolar\ to 0.67~\Msolar\ with a mean value of 0.22~\Msolar. The third panel of Figure~\ref{fig:cdf} displays Arnett \Mni\ CDFs for different SESN sub-types.

\begin{figure}[ht]
\centering
\includegraphics[trim=0.23cm 0.25cm  0.25cm 0.18cm, clip=true, scale=0.53]{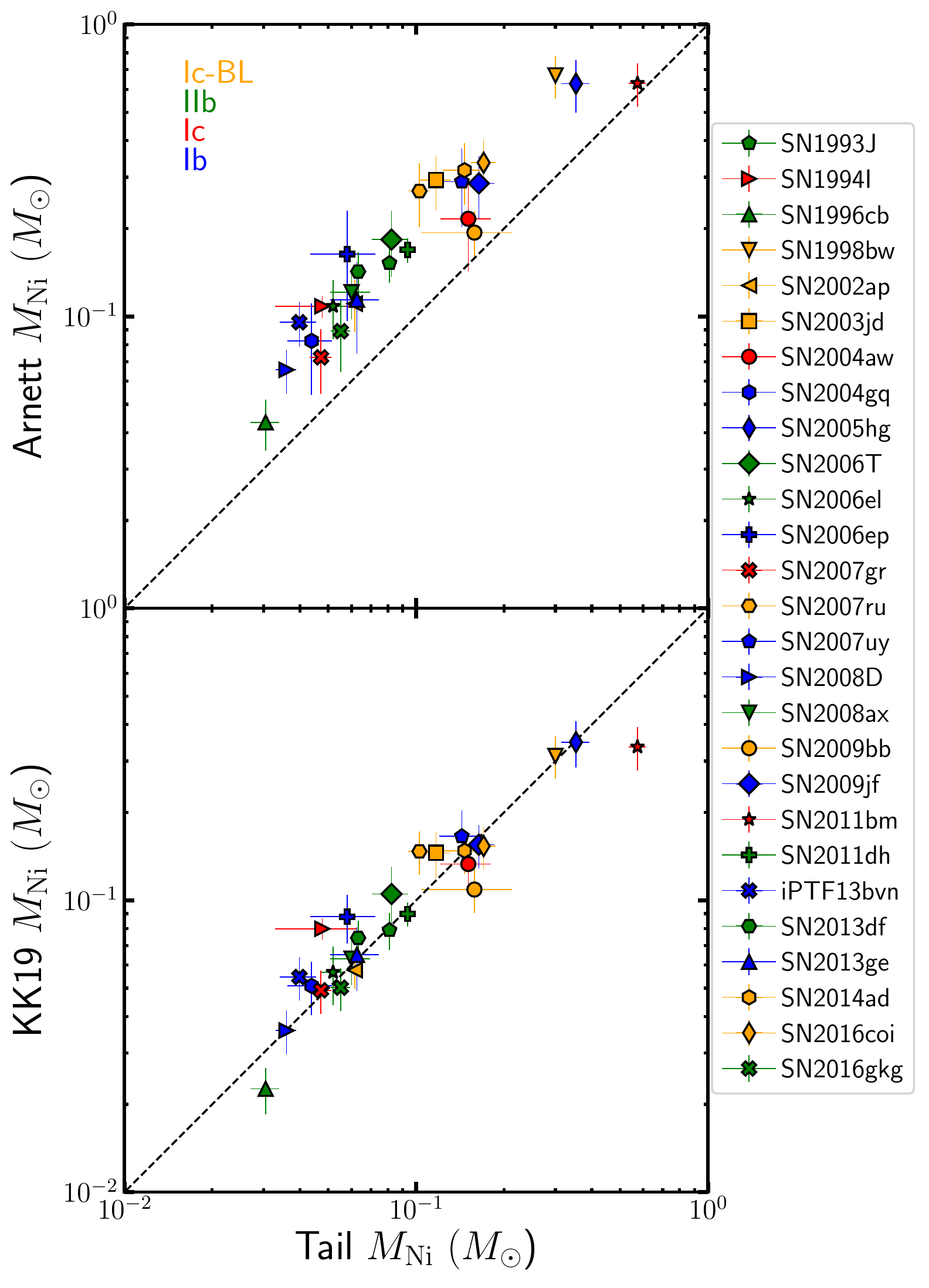}
\caption{\emph{Top:} \Mni\ obtained using the radioactive tail modeling (x-axis) and Arnett's rule (y-axis) for our sample of SESNe. Arnett's rule yields \Mni\ values which are systematically a factor of $\sim$2 larger. \emph{Bottom:} the same but with \Mni\ values calculated using the model of KK19 and our empirically-calibrated $\beta$ values (see \S~\ref{sub:mnikk19}) displayed on y-axis. Substantially better agreement is found. The markers are color-coded to represent the SESN sub-types with yellow, green, red, and blue indicating SNe Ic-BL, IIb, Ic, and Ib, respectively. The dashed black line denotes equality. }
\label{fig:arnettvstail}
\end{figure}

Figure~\ref{fig:arnettvstail} (top panel) presents a comparison between the tail and Arnett \Mni\ for our sample of SESNe. The results highlight the systematic discrepancy between the two methods. The \Mni\ values obtained from the radioactive tail modeling are, on average, a factor of $\sim$2 smaller than those derived using Arnett's rule. The dashed black line indicates the equality condition between both models. Despite the scatter in the severity of this discrepancy for different SNe, the Arnett model consistently overestimates \Mni\ for every SESN in our sample. The overestimation of \Mni\ by Arnett model is also illustrated in Figure~\ref{fig:cdf} (top panel), where the CDF of the Arnett \Mni\ distribution is below that of the tail distribution with a relatively large margin.

The means and standard deviations of our Arnett-derived \Mni\ values for different SN types closely match the values reported in \cite{Lyman2016}, but our median values are lower than those of \cite{Prentice2016} for Type Ib, Ic, and Ic-BL SNe by $\sim0.04$~\Msolar. This discrepancy is primarily due to different approaches in deriving $t_{\rm p}$, which is estimated in \cite{Prentice2016} by measuring the rise time from the half-maximum luminosity to $L_{\rm p}$ (denoted by $t_{-1/2}$) and using a linear empirical correlation for translating $t_{-1/2}$ to $t_{\rm p}$. We also note the Arnett \Mni\ values of \cite{Meza2020} are, on average, 50\% lower than our Arnett estimates. This difference can be traced to their anomalously lower peak luminosities as discussed in \S~\ref{sec:lbol}, above.

The inaccuracy of \Mni\ values obtained from Arnett models has been also shown in several radiative-transfer numerical simulations of SESNe. For example, \cite{Dessart2015, Dessart2016} found that the Arnett's rule overestimated the \Mni\ of SESNe by 50\% and attributed this discrepancy to the fixed electron scattering opacity assumption of Arnett's models. Similarly, \cite{Sukhbold2016} pointed out that Arnett's rule does not hold for their simulations of Type Ib/c SNe evolved from massive single star progenitors. We note that the discrepancy we find between our Arnett and tail \Mni\ values is approximately twice as large as that quoted by \citet{Dessart2016}.

Despite these limitations, Arnett models have been widely used for deriving \Mni\ as well as ejecta masses and kinetic energies of SESNe \citep{Drout2011, Lyman2016, Prentice2016, Prentice2018}. A few observational studies have also indicated contrasts between results from modelling the early and late-time light curves of SESNe. For example, \cite{Valenti2008} evoke a ``two-zone'' model to try to resolve an inconsistency between the explosion parameters derived from early- and late-time light curves of SN2003jd, which they fit with Arnett and radioactive models, respectively. In another study, \citet{Wheeler2015} estimate $T_0$ values for dozens of SESNe by an analytical relation that depends on \Mej\ and kinetic energy, which were rewritten in terms of the observed rise time and photospheric velocity, assuming Arnett's model. The estimated $T_0$ values were found to be in tension with the values measured directly from the light curve tails, which likely is due to the limitations of Arnett's model. More recently, \cite{Meza2020} measured \Mni\ values for a sample of SESNe using a variety of methods. While their tail \Mni\ values are lower limits, they confirm that Arnett values are consistently higher than those derived via other methods. 

For the rest of our analyses, we assume that our tail \Mni\ estimates are more realistic than those of the Arnett's model. This is because the ejecta are expected to be transparent to optical photons over the tail. Therefore, the bolometric luminosity traces the instantaneous heating rate without any further assumption regarding the self-similarity of the energy density profile, which is a fundamental assumption in the Arnett-like models (KK19).

\subsection{Comparison of Stripped-Envelope and Type II SN \Mni}
\label{sub:compTypIIvsSESNe}
As described \S~\ref{sec:intro}, by comparing measurements of \Mni\ for 115 H-rich Type II SNe and 145 SESNe previously published in the literature, \citet{Anderson2019} identified a discrepancy in their distributions, with SESNe displaying a mean \Mni\ that was a factor of $\sim$6 larger than their H-rich counterparts. Subsequently, \citet{Meza2020} computed \Mni\ directly for a smaller sample of SESNe, using a number of methods (Arnett's rule, KK19, and radioactive tail modeling) in a uniform manner, demonstrating that a statistically significant discrepancy remains. However, the tail \Mni\ values presented in \citet{Meza2020} are strictly lower limits to the true \Mni, as they assume complete trapping of $\gamma$-rays, while the Arnett and KK19-based values may each contain systematic biases (in the latter case because they adopt $\beta$ values which have yet to be observationally calibrated; see \S~\ref{sub:khatami}). Thus, while sufficient to robustly demonstrate that SESN have a different \Mni\ distribution than Type II SNe, the magnitude of this discrepancy remains somewhat uncertain. 

\begin{figure}[ht!]
\centering
\includegraphics[trim=0.2cm 0.2cm  0.2cm 0.2cm, clip=true, scale=0.64]{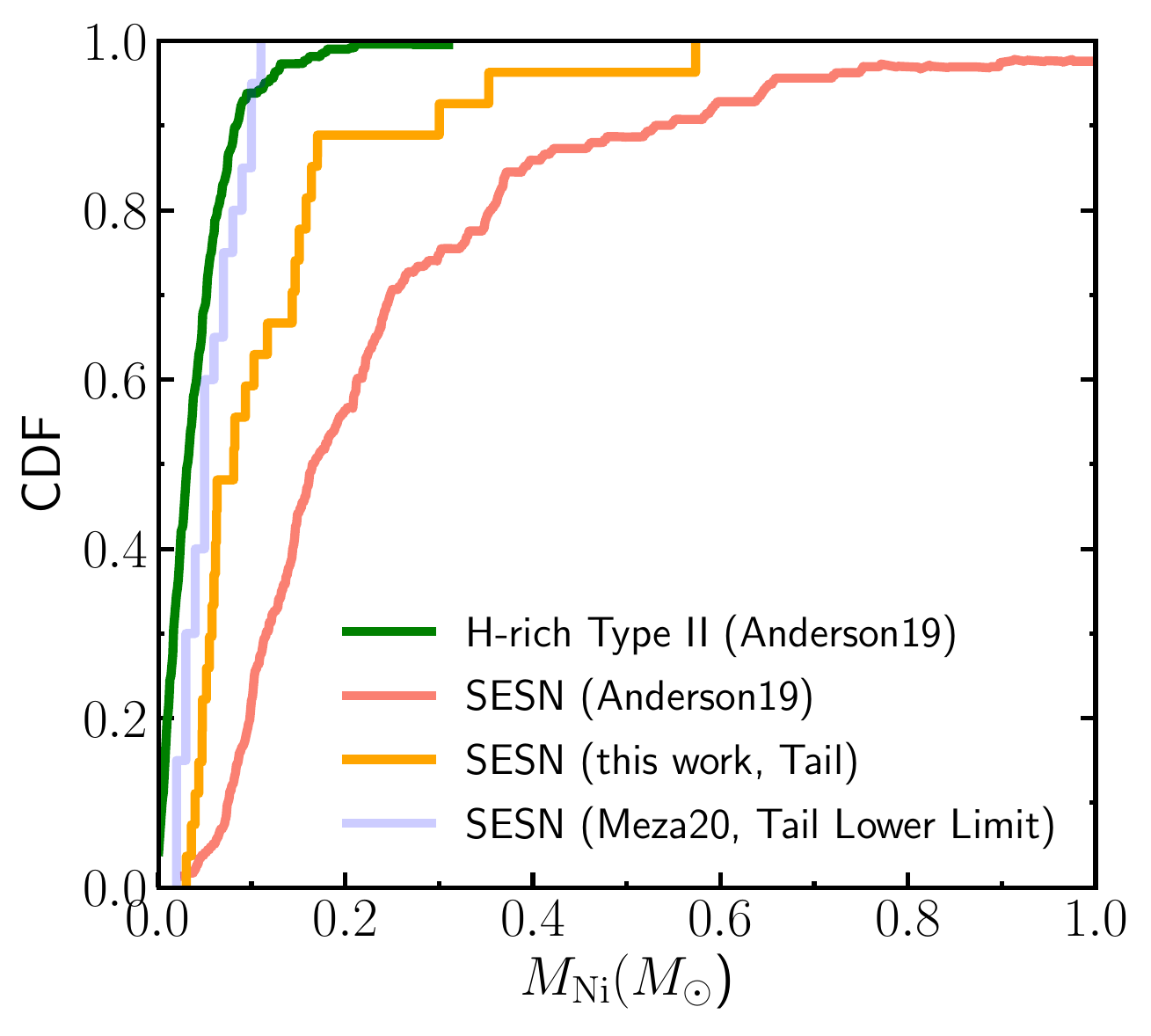}
\caption{Cumulative Distribution Functions of \Mni\ for SESN vs.\ H-rich Type II SNe. The green curve represents tail-based \Mni\ values for Type II SNe compiled \citet{Anderson2019}, while red, light blue and orange curves represent three different distributions for SESNe: the primarily Arnett-based \Mni\ values compiled from the literature by \citet{Anderson2019}, lower limits on tail \Mni\ computed by \citet{Meza2020}, and the tail \Mni\ values derived in \S~\ref{sec:res}. The mean value of the tail-based \Mni\ values found in this work are a factor of $\sim$3 higher than the distribution of Type II SNe.}
\label{fig:anderson_cdf}
\end{figure}

Here, we compare the CDF of \Mni\ for SESNe derived from our radioactive tail measurements to that for the 115 H-rich Type II SNe from \citet{Anderson2019}. \Mni\ for most of this sample of Type II SNe were calculated using the bolometric luminosity of the radioactive tail of the light curve assuming full trapping of the $\gamma$-rays. Figure~\ref{fig:anderson_cdf} illustrates the \Mni\ CDF of our sample of SESNe (orange curve) and the H-rich Type II SNe (green curve). For reference, we also show the original sample \Mni\ values of SESNe compiled from the literature by \citet[pink curve]{Anderson2019}---most of which were obtained using Arnett's rule---and the lower limits of \Mni\ from \citet[light blue curve]{Meza2020}. We see that our distribution of \Mni\ for SESNe measured from the radioactive tail lies between that of the Arnett-based values of \citet{Anderson2019} and the tail upper limits of \citet{Meza2020}, as expected.

Overall we find that our sample of SESNe have a mean value ($\sim$0.12 \Msolar; Table~\ref{tab:mni_stat}) which is a factor of $\sim$3 larger than that of H-rich Type II SNe ($\sim$0.044 \Msolar). This is a factor of $\sim$2 smaller than the initial discrepancy reported by \citet{Anderson2019} based on Arnett measurements.  We conduct K-S tests to the \Mni\ CDFs of H-rich Type II SNe \citep{Anderson2019} and SESNe in this work. The test gives D-value~$=0.52$ and p-value~$=10^{-7}$, meaning that the CDFs are inconsistent with being drawn from the same distribution. By excluding Type Ic-BL SNe from the test, we find D-value~$=0.49$ and p-value~$=5\times 10^{-5}$, which similarly confirms that Type IIb/Ib/c SNe and H-rich Type II SNe are inconsistent with being drawn from the same distributions.

\begin{figure}[ht]
\centering
\includegraphics[trim=0.2cm 0.2cm  0.2cm 0.2cm, clip=true, scale=0.54]{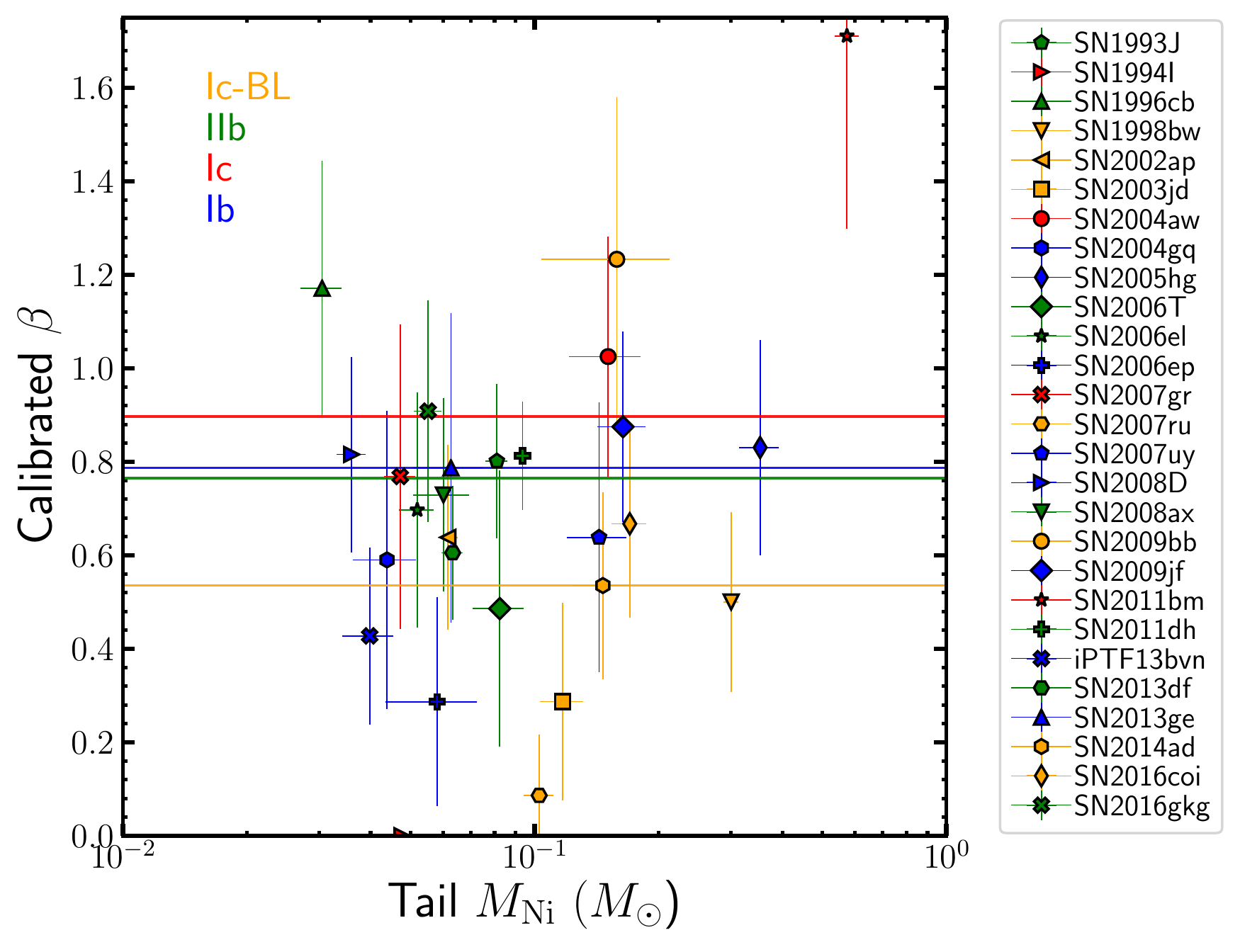}
\caption{Values for the $\beta$ parameter given in Equation~\ref{eq:khatami} for the SESNe in our sample, calculated using the observed $t_{\rm p}$, $L_{\rm p}$, and tail-based \Mni\ as inputs. The results are color-coded based the SN sub-type. $\beta$ is a dimensionless parameter defined by KK19 and is correlated with different physical effects such as composition, asymmetries, and the radial extent of \nickel within ejecta. The horizontal lines indicate the median $\beta$ value Type IIb (green), Type Ib (green), Type Ic (red), and Type Ic-BL (yellow) SNe.}
\label{fig:tunedbeta}
\end{figure}

As in \citet{Meza2020} we find that a majority of this discrepancy come from the lack of SESNe in our sample with low \Mni\ values. The lowest tail \Mni\ in our sample is 0.03 \Msolar, while an incredible $\sim$48\% of Type II SN have \Mni\ lower than this value. If we recompute the K-S tests described above, but considering only Type II SN with \Mni$>$0.03\Msolar, we find a p-value$=$0.008 for the full sample of SESNe and p-value$=$0.06 when Type Ic-BL are excluded. This indicates that the sample of IIb/Ib/Ic SNe are marginally consistent with being drawn from the same population as the high \Mni\ Type II SNe.

\subsection{Calibration of $\beta$ values from KK19}
\label{sub:khatami}
In \S~\ref{sub:comp} we demonstrated that Arnett-based models yield \Mni\ values that are a factor of 2 larger than those found from modelling the radioactive tail. While obtaining tail-based \Mni\ measurements for all SESNe would be ideal, in practice the requisite photometric data exists for only a subset of events. Thus, another means to estimate \Mni\ from photospheric data alone would be beneficial.

\begin{deluxetable*}{l |ccc|ccc|ccc|ccc|ccc}
\tabletypesize{\footnotesize}
\tablecolumns{16}
\tablewidth{0pc}
\tablecaption{Table of Derived Parameter Statistics}
\tablehead{\colhead{}& \multicolumn{3}{|c|}{$\beta^{*}$} &\multicolumn{3}{c|}{$t_{\rm p}$ (days)} &\multicolumn{3}{c|}{Log $L_{\rm p}$  (erg~s$^{-1}$)}  & \multicolumn{3}{c|}{$f^{\dagger}$}  &
\multicolumn{3}{c}{Log $f L_{\rm p}$ (erg~s$^{-1}$)}  \\
{SN Type} & {Mean} & {Median} & {Std} &  {Mean} & {Median} & {Std} & {Mean} & {Median} & {Std} & {Mean} & {Median} & {Std} & {Mean} & {Median} & {Std}
}
\startdata
IIb &  0.78 & 0.77 & 0.19 & 20.2 & 21.5 & 2.5 & 42.37 & 42.34 & 0.21 & 0.24 & 0.26 & 0.12 & 41.87 & 41.78 & 0.21\\
Ib & 0.66 & 0.79 & 0.21 & 17.4 & 16.8 & 3.0 & 42.61 & 42.39 & 0.30 & 0.29 & 0.24 & 0.10 &  42.03 & 41.94 & 0.29  \\
Ic & 0.88 & 0.90 & 0.61 & 17.0 & 12.7 & 10.5 & 42.67 & 42.67 & 0.18 & 0.10 & 0.13 & 0.29 & 41.97& 41.62 &0.32 \\
Ic-BL & 0.56 & 0.54 & 0.33 & 14.4 & 15.0 & 2.5 & 42.87 & 42.84 & 0.19 & 0.32 & 0.36 & 0.17 & 42.49 & 42.47 & 0.27\\
All & 0.70 & 0.70 & 0.34 & 17.4 & 16.6 & 5.2 & 42.67 & 42.55 & 0.30 & 0.26 & 0.29 & 0.18 & 42.17& 41.94 & 0.36  \\
\enddata
\tablenotetext{*}{The parameter $\beta$ is discussed in \S~\ref{sec:nickel} and obtained in \S~\ref{sub:khatami}.}
\tablenotetext{\dagger}{The excess power factor $f$ is defined in \S~\ref{sub:additional_power}.}
\label{tab:beta_stat}
\vspace{-0.2in}
\end{deluxetable*} 

As discussed in \S~\ref{sec:nickel}, KK19 proposed an analytical model that relates the peak luminosity and its epoch to a general heating function without relying on some of the simplifying assumptions adopted by Arnett's models.  This new model, described in Equation~\ref{eq:khatami}, depends on a dimensionless parameter $\beta$ in addition to \Mni, $t_{\rm p}$ and $L_{\rm p}$. KK19 suggested $\beta=9/8$ for Type Ib/c SNe based on the radiative transfer simulations of SESN light curves from \cite{Dessart2016} and $\beta=0.82$ for Type IIb/pec SNe based on the observed light curve of SN1987A. However, numerical simulations may not fully represent the behavior of real SESN SNe and the light curve of SN1987A is very different than that of Type IIb SNe.  More reliable constraints on $\beta$ can be obtained from the observed sample of SESNe with independent \Mni\ values measured from their radioactive tails. 

We use Equation~\ref{eq:khatami} to calculate the value of $\beta$ inferred for each SESN in our sample given their tail \Mni, $t_{\rm p}$ and $L_{\rm p}$ provided in Table~\ref{tab:sn_derived_param}. The uncertainty in the derived $\beta$ values has contributions from the error in tail \Mni, $L_{\rm p}$, and $t_{\rm p}$. In Figure~\ref{fig:tunedbeta} we plot the derived $\beta$ values versus tail nickel mass for individual SNe. The results exhibit a significant scatter, with $\beta$ ranging from $\approx$ 0.0 to 1.7 and a mean value of 0.70. This is lower, on average, than the $\beta=9/8$ suggested by KK19 for SESN based on the models of \cite{Dessart2016}. This comparison will be examined in more detail in \S~\ref{sub:diffmodels}.  The horizontal lines in Figure~\ref{fig:tunedbeta} indicate the median $\beta$ value for each SESN sub-type. Table 4 provides summary information on the mean, median and standard deviation of the derived $\beta$ values for each SN sub-type. The median values of $\beta$ for Type IIb, Ib, and Ic SNe are roughly similar, but the standard deviation of Type Ic SN is a factor $\sim$3 larger due to the small sample size of objects of this type. Type Ic-BL SNe has the smallest median $\beta$ value of 0.54, which is $\approx$ 30\% smaller than that of other SESN types.

Two SNe in our sample, SN1994I and SN2007ru, have $\beta$ close to zero, which may suggest that the derived tail \Mni\ is inadequate to produce the observed peak luminosity, $L_{\rm p}$. Recall that for a fixed \Mni\ and $t_{\rm p}$, a lower $\beta$ value will yield a higher peak luminosity (see \S~\ref{sec:KK19model}). Both of these SNe are fast declining and will be discussed further in \S~\ref{sec:disc}.

\subsection{Improved Photospheric \Mni\ Estimates from the Median Calibrated $\beta$ Values}
\label{sub:mnikk19}

Using the results from \S~\ref{sub:khatami} we now assess whether, in practice, the model of KK19 can be used to obtain more reliable \Mni\ estimates than Arnett for SESNe from photospheric data alone. In Table~\ref{tab:sn_derived_param} we list \Mni\ values for each SN that have been calculated using their observed $L_{\rm p}$ and $t_{\rm p}$ in conjunction with the median $\beta$ value for each SN sub-type listed in Table~\ref{tab:beta_stat}. Errors listed in Table~\ref{tab:sn_derived_param} account only for the errors in $L_{\rm p}$ and $t_{\rm p}$ and do include the impact of the standard deviation in the distribution of $\beta$ values.
Despite the significant scatter in $\beta$ values found for individual SNe, we find that the procedure of calculating \Mni\ assuming the KK19 model and the median $\beta$ value for each SN sub-type offers a significant improvement over Arnett-based measurements, both for the overall distribution of \Mni\ values for SESN \emph{and} for individual objects. These two effects are demonstrated in Figures~\ref{fig:cdf} and ~\ref{fig:arnettvstail}, respectively. 

In the top panel of Figure~\ref{fig:cdf} we present the CDF of \Mni\ calculated using KK19 with the median calibrated $\beta$ in comparison to that of the radioactive tail and Arnett methods. As shown, not only is the mean value of the KK19 \Mni\ distribution the same as that from the radioactive tail measured \Mni\ (shown by a vertical lines; see also Table~\ref{tab:mni_stat}), but the overall CDF of KK19-measured \Mni\ values closely approximates that of the radioactive tail \Mni\ measurements. In the bottom panel of Figure~\ref{fig:cdf} we also display the CDFs of KK19 \Mni\ estimates separated by SN sub-type. As listed in Table~\ref{tab:mni_stat}, the median values of these distributions are all within $\approx$10\% of those calculated from the radioactive tail. 

In Figure~\ref{fig:arnettvstail} we demonstrate that this agreement extends to individual objects. The KK19-measured \Mni\ values for each SNe (bottom panel) are much closer to their tail counterparts compared to the Arnett-derived ones (top panel). We find that, on average, the KK19 \Mni\ values are within $\sim$17\% of the tail-derived values. The largest variations occur for the Type Ic SN\,1994I and SN\,2011bm for which the nickel mass is overestimated by $\sim$65\% and underestimated by $\sim$40\%, respectively. In contrast, the Arnett-based models systematically over predict \Mni\ by a factor of $\sim$2 (100\%) compared to the tail-derived values. We therefore conclude that in cases where the radioactive tail is not observed, the KK19 model with median calibrated $\beta$ values listed in Table~\ref{tab:beta_stat} should be used to calculate \Mni\ for SESNe.

\subsection{Inferred $\beta$ Values for Theoretical SESN Light Curves}
\label{sub:diffmodels}

In addition to providing improved estimates for \Mni\ from photospheric data alone, the $\beta$ values calculated for our observed SESNe encode information on the explosion properties and progenitors of the population. Effects such as composition, asymmetry and additional power sources will impact the degree to which the internal energy of the ejecta lags or leads the observed luminosity at the time of peak. In order to assess if the observed population of SESN matches expectations from theory, and to gain insight into the physical processes that dictate $\beta$ in observed events, we calculate the $\beta$ values for a set of analytical and numerical light curves models available in the literature.

\subsubsection{Arnett Models}

First, as a baseline, we derive $\beta$ for a grid of light curves calculated using an analytic Arnett model.  We take $t_{\rm p}$ and \Mni\ in the ranges 5--40~days and 0.02--0.5~\Msolar, respectively, which correspond to the approximate ranges found for our observed sample . Given a pair of $t_{\rm p}$ and \Mni, Arnett's rule gives $L_{\rm p}$; we then compute $\beta$ from Equation~\ref{eq:khatami} using $t_{\rm p}$, $L_{\rm p}$, and \Mni. The resulting values of $\beta$ for this grid are plotted as a grey shaded region in Figure~\ref{fig:models}. As expected given the inconsistency between Arnett and tail-derived nickel masses shown above, these $\beta$ values are inconsistent with those of our observed population. In particular, the Arnett models occupy a parameter space with high $\beta$ values in the range of 1.55--1.95, while all but one observed SESN (SN\,2011bm) has a calibrated $\beta$ $<$ 1.25.

\subsubsection{Large Grids of Numerical SESN}

Next, we calculate the implied $\beta$ values for two large suites of simulated SESN light curves from \cite{Dessart2016} and \cite{Ertl2019}. Both sets of models consider the explosion of a grid H-poor stars, but utilize different pre-SN stellar structures, explosion assumptions, and hydrodynamic/radiative transfer codes. \cite{Dessart2016} consider a subset of the SN progenitors stripped via close binary interaction that were evolved in \citet{Yoon2010}. These models have final pre-explosion masses between 3.0 and 6.5 M$_\odot$ (initial masses between 16 and 60 M$_\odot$) and final compositions chosen to span the range of SESN sub-types: Type IIb (defined as $>$ 50\% He plus some residual H in the outer envelope), Type Ib ($\approx$35 \% He), and Type Ic (H and He deficient). \citet{Dessart2016} uses a piston in order to produce four different explosion energies for each pre-SN structure and also consider two different levels of mixing of radioactive materials. The purpose was to investigate how these physical properties map onto observables, rather than ascertaining what explosion energy and 3D effects would be achieved for a given pre-SN structure a priori. Final SN light curves were calculated with the 1D non-local thermodynamic equilibrium (non-LTE) radiative-transfer code CMFGEN \citep{Dessart2010}, and thus account for time and wavelength-dependent opacity variations.

\begin{figure}[t!]
\centering
\includegraphics[trim=0.2cm 0.2cm  0.5cm 1cm, clip=true, scale=0.6]{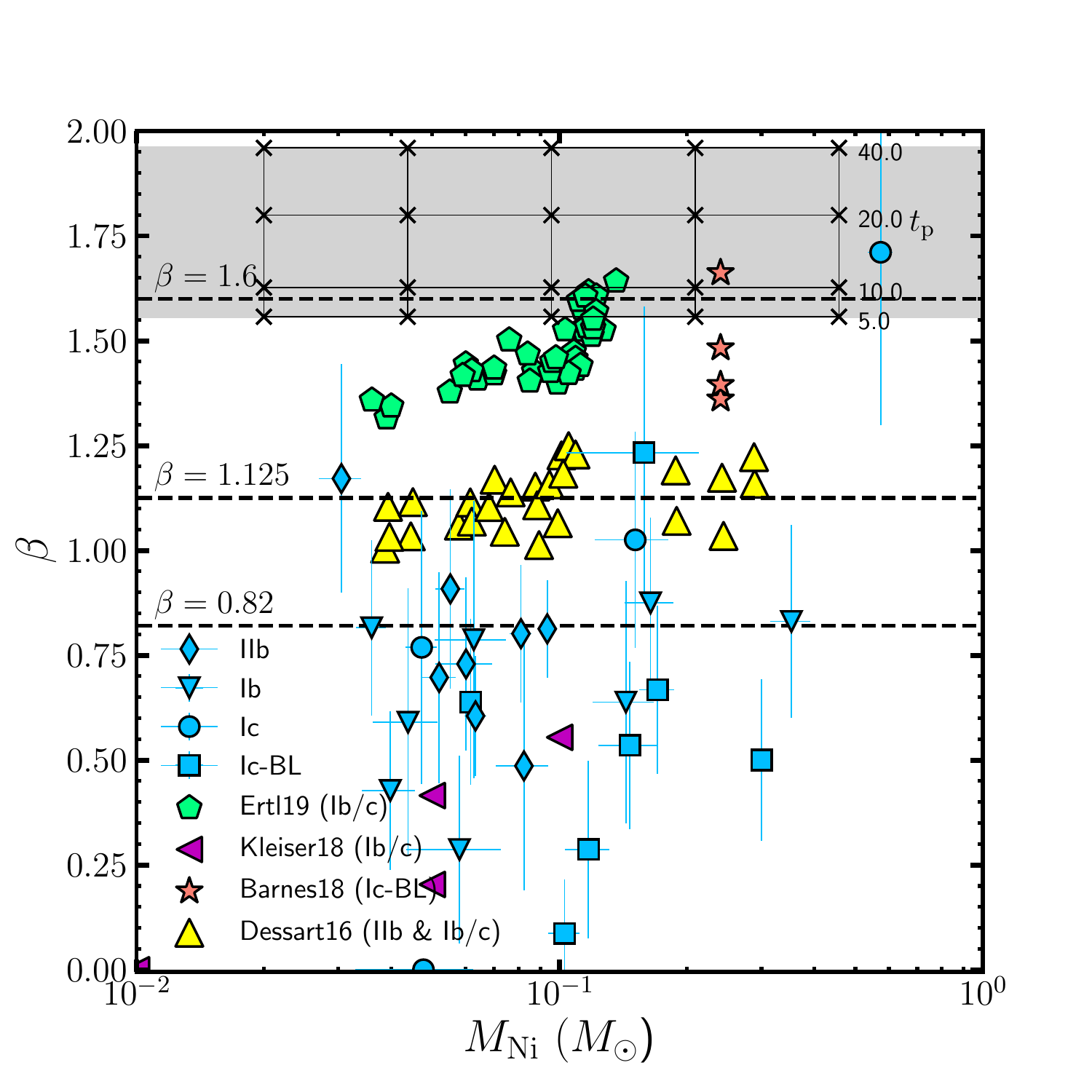}
\caption{Values for the $\beta$ parameter given in Equation~\ref{eq:khatami} found for the observed sample of SESNe in comparison to those calculated from a variety of theoretical models. The gray region represents the parameter space that Arnett light curves take for various pairs of $t_{\rm p}$ and \Mni. The numerical models of  \cite{Dessart2016}, \cite{Barnes2018}, \cite{Kleiser2018b,Kleiser2018a}, and \cite{Ertl2019} are shown with yellow, red, magenta, and green markers, respectively. The blue markers denote $\beta$ obtained for our sample of SESNe. The  diamond, inverted triangle, circle, and square blue markers represent SN types IIb, Ib, Ic, and Ic-BL, respectively. Overall, the observed population of SESNe possess lower $\beta$ values with more scatter than the numerical models. Only the models of \citet{Kleiser2018b,Kleiser2018a}, in which shock cooling contributes a significant fraction of the peak luminosity, overlap with the bulk of the observed population. The horizontal dashed lines represent $\beta=\{0.82,1.125,1.6\}$ suggested by KK19 for Type IIb, Ib/c, and Type Ia SNe, respectively.  }
\label{fig:models}
\end{figure}

In contrast, \citet{Ertl2019} consider the explosion of the He star models of \cite{Woosley2019}, which are assumed to have lost their H envelopes due to binary interactions prior to the onset of He-ignition, and are evolved to core-collapse in the KEPLER hydrodynamic code \citep{Weaver1978}. Thus, all pre-SN models are H-deficient, but likely lead to a combination of Type Ib and Type Ic SN, with final surface He mass fractions spanning 0.16 to 0.99.
Unlike in \citet{Dessart2016} the SN explosions are carried out in a neutrino-hydrodynamics code P-HOTB \citep{Janka1996}, giving constraints on explosion energies, nickel masses, and remnant masses. The progenitor models that lead to a successful SN have initial He star masses in the range 3.3--19.75~\Msolar, which roughly translates to ZAMS mass range of 16--51~\Msolar. For these events, bolometric light curves are calculated by post-processing the P-HOTB results in KEPLER. While KEPLER treats electron scattering directly, a constant additive opacity must be adopted to account for the effects of atomic lines. \citet{Ertl2019} chose this `line' opacity to match that of SESN near peak. 

For each light curve published in \citet{Dessart2016} and \citet{Ertl2019}, we compute $\beta$ from the published \Mni, $t_{\rm p}$, and $L_{\rm p}$ and Equation~\ref{eq:khatami}. The results are presented in Figure~\ref{fig:models}. The $\beta$ of 
\citeauthor{Ertl2019}'s models (green) span then range 1.31--1.64 with more massive initial He stars having relatively higher nickel masses and $\beta$ values. These are larger than the $\beta$ values found for all of the of models of \citeauthor{Dessart2016}, which are in the range 1.00--1.25 with a mean value of 1.12. Interestingly, the $\beta $ values of the observed SESNe are considerably smaller and the scatter in the observed $\beta$ values much larger than those of either set of numerical models. In addition, while the observed SESNe span a similar range of \Mni\ as the models of \citet{Dessart2016}, $\sim$33\% have tail-based \Mni\ higher than any any of the models of \citet{Ertl2019}, which were designed to self-consistently determine the radioactive material that can be synthesized by neutrino-driven explosions. We discuss the implications of these results in \S~\ref{sec:disc}. 
For comparison, in Figure~\ref{fig:models} we also mark the values $\beta=\{0.82,1.125,1.6\}$ that are recommended for Type IIb, Ib/c, and Ia SNe, respectively, by KK19. While $\beta = 0.82$ only slightly overestimates the mean $\beta$ value of 0.78 obtained for the observed Type IIb SNe, $\beta=1.125$ substantially overestimates the mean $\beta$ values for Type Ib and Ic SNe (see Table~\ref{tab:beta_stat}).

\subsubsection{Specialized SESN Models}\label{sub:specialized}

Finally, we also examine the light curve models from two specialized models, which were each designed to probe a specific physical effect that may be present in SESN. \cite{Barnes2018} performs a 2D relativistic hydrodynamic simulation with radiative transport in order to model a single jet-driven explosion. They adopt an analytic pre-SN model with a mass of 3.9~\Msolar\ and inject an engine with an engine of $\sim$2$\times$10$^{52}$ ergs. The resulting explosion would be classified as a Type Ic-BL, and synthesizes 0.24 M$_\odot$ of $^{56}$Ni. By modelling in multiple dimensions \citet{Barnes2018} find that both the ejecta density profile and distribution of radioactive material are aspherical, and generate light curves for different viewing angles. We compute the $\beta$ that would be inferred from each of these angles and plot the results as red stars in Figure~\ref{fig:models}. We find $\beta$ in the range 1.35--1.65 with models viewed from directions more aligned along the polar axis having progressively higher $\beta$. These results lie close to those of the observed Type Ic-BL SN\,2009bb ($\beta$ $=$1.23; \Mni\ $=$ 0.158 M$_\odot$), but yield significantly larger $\beta$ values than observed for most of the Type Ic-BL in our sample ($\langle \beta \rangle = 0.56$; see Table~\ref{tab:beta_stat}).

\citet{Kleiser2018b,Kleiser2018a} examine the explosion of H-free stars which have either been inflated to large radii or are embedded in a CSM shell ejected shortly before explosion. In both cases, the effective pre-SN radius can be large ($\gtrsim$30 R$_\odot$) and the subsequent cooling of shock deposited energy can lead to substantial luminosity beyond that provided by $^{56}$Ni. Using a combination of the MESA stellar evolution code, hydrodynamic simulations, and the Sedona radiative transport code \citet{Kleiser2018b,Kleiser2018a} model the light curves that would result from the explosion of such systems, finding luminosities of $\log L =$ 41.2--42.5 on timescales of 10$-$20 days. Originally proposed as a means to explain the class of rapidly evolving Type I SN (e.g. SN2010X; \citealt{Kasliwal2010}), a majority of the models of \citet{Kleiser2018b,Kleiser2018a} are computed without contributions from $^{56}$Ni. However, \citet{Kleiser2018b} also provide six fiducial models in which they add 0.01, 0.05, and 0.1 M$_\odot$ of $^{56}$Ni to one of their models with two levels of mixing. 
We calculate $\beta$ for the three ``strongly'' mixed models---which yield relatively smooth, as opposed to strongly double-peaked, light curve morphologies most similar to observed Type Ib/c SN---and plot the results in Figure~\ref{fig:models} (magenta triangles). We find $\beta$ values of $\sim$0.0--0.55, which overlap with the \emph{lower} end of observed SESNe. Implications of these results are discussed below.

\begin{figure*}[ht]
\centering
\includegraphics[trim=2.2cm 2.5cm 2cm 3.5cm, clip=true, scale=0.55]{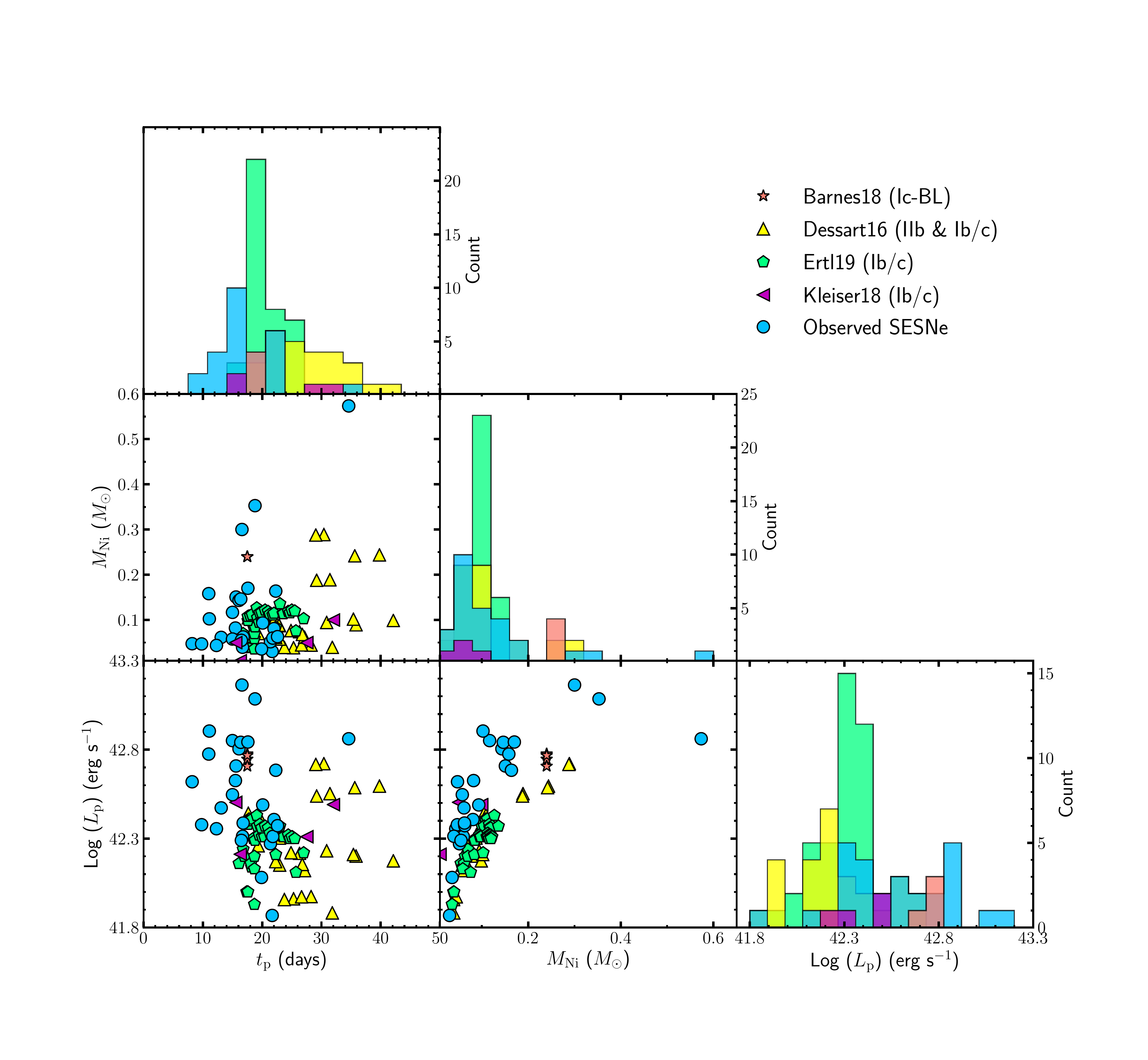}
\caption{Pairwise relationship of $\log L_{\rm p}$, $t_{\rm p}$, and \Mni\ for numerical models of  \cite{Dessart2016} (yellow upper triangles), \cite{Barnes2018} (red stars), \cite{Kleiser2018b,Kleiser2018a} (magenta left triangles),  \cite{Ertl2019} (green pentagons) and our sample of SESNe (blue circles). Diagonal panels display the histogram of the three parameters. \Mni\ for the observed sample are those derived from modelling the late-time tail (\S~\ref{sub:tail}). Overall the observed population of SESNe display shorter rise times and have substantially higher peak luminosities for a given \Mni\ than the numerical models.}
\label{fig:modelparam}
\end{figure*}

\section{Discussion} \label{sec:disc}

In the sections above, we calculated \nickel masses for 27 SESNe based on their late-time tails. We confirm that these masses are systematically lower than those derived by Arnett-like analytical models. These masses allow us to \emph{observationally} calibrate the $\beta$ value introduced in KK19 based on their observed rise times and peak luminosities. Despite scatter, we demonstrate that calculating \Mni\ using the medians of our empirically calibrated $\beta$ values offers a significantly improved estimation when only photospheric light curve data is available. However, in doing so we find that (a) the $\beta$ values inferred for SESNe are systematically lower than those found from most numerical simulations of SESN explosions and (b) a systematic discrepancy remains between the \nickel masses for SESNe and Type II core-collapse SNe. In the sections below, we discuss the possible origins for each of these discrepancies, and their implications for the progenitors and explosion mechanism of stripped envelope core-collapse SN. 

\subsection{Possible Origins of Low $\beta$ Values}\label{sub:lowbeta}
In \S~\ref{sub:diffmodels}, we presented the $\beta$ values inferred for different numerical models of SESNe. The results show that most numerical models give higher $\beta$ values compared to observations. In order to  disentangle the origin of this discrepancy, in Figure~\ref{fig:modelparam} we present the pairwise dependence and histograms of the three quantities that determine $\beta$, i.e., $L_{\rm p}$, $t_{\rm p}$, and \Mni, for both the numerical models detailed in \S~\ref{sub:diffmodels} and observed SESNe. This figure illustrates that: 
\begin{enumerate}
    \item While both the observed sample and model SESN show a correlation between \Mni\ and $L_{\rm p}$, the observed objects exhibit considerably ($\sim$0.3--0.4 dex) larger peak luminosities for a given \Mni.
    \item The observed SESNe tend to have shorter rise times compared to the models. The median rise time for the entire sample is 16.6 days, as compared to 19.5 and 26.8 days for the \citet{Ertl2019} and \citet{Dessart2016} models, respectively.
\end{enumerate}

As described in \S~\ref{sec:KK19model} both rise time and peak luminosity are \emph{inversely} proportional to $\beta$. Thus, shorter rise times would primarily act to \emph{increase} $\beta$. Indeed, it appears that the main driver between the different $\beta$ values of the \citet{Ertl2019} and \citet{Dessart2016} models is that \citeauthor{Ertl2019} find shorter rise times for a given \Mni. This effect was discussed in \citet{Ertl2019} and is primarily due to their adoption of a constant line opacity. In contrast, the higher luminosity for a given \Mni\ would act to lower $\beta$, and this is therefore likely the origin of the discrepancy in $\beta$ displayed in Figure~\ref{fig:models}. Here we investigate the possible physical origins of this discrepancy, as well as the scatter in effective $\beta$ displayed by the observed sample.

\subsubsection{Systematic Effects Due to Distance and Reddening}

{As outlined in \S\ref{sec:distance} and \S\ref{sec:extinction} there are systematic uncertainties in our distances and extinction values due to our choice of cosmological parameters and $R_V$ value, respectively. While these will lead to systematic uncertainties on our bolometric luminosities and \nickel masses, we find that they, alone, cannot explain the discrepancy in $\beta$ values described above. In particular, an $\sim$8\% increase distances due to adoption of the Plank cosmological parameters would lead to an $\sim$17\% increase in \emph{both} the measured peak luminosity and inferred tail \Mni\ show in Figure~\ref{fig:modelparam} (the latter due to the linear relationship between tail luminosity and \nickel mass). For a fixed rise time, if both \Mni\ and $L_{\rm p}$ increase by the same fraction, then the $\beta$ value inferred from Equation~\ref{eq:khatami} will be unchanged.} 

{Similar arguments apply for reddening corrections: while variations in $R_V$ values lead to different extinction corrections in the bands considered, the impact of this is muted because it would impact the light curve at \emph{both} at peak and on the tail. While the effect does not completely cancel as it does for distances (due to the color dependence of the bolometric corrections derived in \S\ref{sec:lbol}), we find that adopting an $R_V$ value of 2.1--4.1 would only impact our calibrated $\beta$ values by $\lesssim$4-10\%, which is less than the errors on the $\beta$ values themselves. Thus, systematic uncertainties due to distances and reddening cannot, by themselves, explain either the low $\beta$ values inferred for the observed sample of SESNe, or the conclusion that the observed sample displays higher peak luminosities for a given \Mni. We therefore investigate other possible explanations in the sections below.} 



\subsubsection{Dark Period}\label{sec:dark}
The light curves of some SESNe are expected to have a ``dark period'' between the explosion epoch and the first observable light if they lack prominent cooling envelope emission and their \nickel is deposited  deep within the ejecta \citep{Piro2013}. This dark period is roughly the time that takes for the diffusion front to move inward (in a Lagrangian sense) and reach the shallowest regions of the ejecta that contain \nickelwospace. KK19 use numerical simulations to show that for a completely central heating source, the dark period could be as large as 20~days, while the models of \cite{Dessart2016} typically have dark periods $\lesssim$5 days. 

While the explosion epochs for many of the Type IIb SN in our sample were determined by the presence of cooling envelope emission, it is possible other events may possess a non-negligible dark period.
This would have two primary effects on our analysis: (i) our current rise times would be underestimated as the explosion would occur earlier than a power-law fit to the light curves implies and (ii) our current tail-based nickel masses would be underestimated as the $^{56}$Ni $\rightarrow$ $^{56}$Co $\rightarrow$ $^{56}$Fe decay chain would need to reproduce the same tail luminosity at a longer time post-explosion. 
Both effects could alleviate some of the discrepancies between the observed and model SESN in Figure~\ref{fig:modelparam}.

To quantify the effect of a possible dark period on our analysis in \S~\ref{sec:res}, we increase the rise time by 5 days and recompute both tail \Mni\ and $\beta$ for our sample of SESNe. The results show that, on average, the tail \Mni\ values would increase by $\sim$14\%, while the inferred $\beta$ values would actually further \emph{decrease} by $\sim$8\% compared to those listed in Table~\ref{tab:sn_derived_param}. We therefore conclude that while a dark period could explain some of the discrepancy between the rise times of observed SESNe and numerical models, the subsequent increase in inferred \Mni\ is insufficient to offset this effect, and an even larger discrepancy between the $\beta$ of the observed SESNe and numerical models would result. 

\subsubsection{Composition, Opacity, and Recombination}
\label{sub:composition}

Composition can influence the morphology of SESN light curves, primarily through its influence on the opacity of the ejecta. In practice, the opacity will also depend both on the age of the SN and the spatial location of of the diffusion front, as effects such as ionization, recombination, and line blanketing will modify the opacity compared to constant pure electron scattering. Notably, while ionized, the opacity of a given material will be dominated by electron scattering and will subsequently fall to significantly once recombined (e.g. KK19, \citealt{Piro2014a}). KK19 investigate the impact of ion recombination on light curve rise times, peak luminosities and $\beta$ values for ejecta with varying compositions. For ejecta with higher recombination temperatures, the opacity will fall at an earlier time. This leads to a shorter dark period, shorter observed rise time, and higher peak luminosity, which combine to yield a \emph{lower} $\beta$ value. KK19 find that for a central heating source and ejecta with recombination temperatures of $\sim$12000 K, 6000 K, and 4000 K (appropriate for He, H, and C/O compositions, respectively), $\beta$ values of 0.70, 0.94, and 1.12 result. $\beta$ $=$ 0.7 also corresponds to the mean value found for our observed SESN sample (see Table~\ref{tab:beta_stat}). Thus, it may be possible to reconcile all of the (i) short rise times (ii) high peak luminosities and (iii) low $\beta$ values of the observed SESN sample if the $^{56}$Ni is primarily diffusing through \emph{He-rich} ejecta. However, we note that the models \citet{Dessart2016} which include full non-LTE radiation transport and wavelength dependent opacities, all display $\beta$ around 1.12, despite the fact that over 60\% of their models have compositions that are $>$50\% He. This implies that the $^{56}$Ni synthesized in the explosions is primarily diffusing through the denser CO cores, whose opacities remains high long after the He envelopes. In this case, the surface He would have recombined at earlier times and would be effectively transparent near maximum light, consistent with the low blackbody temperatures observed for many SESN near maximum ($\sim$9000 K; \citealt{Piro2014a}). 

Thus, given that all SESN progenitors will possess a CO core, the effects of He recombination can likely only explain the low $\beta$ values of observed SESN if a significant fraction of their \Mni\ is mixed out into a He-rich envelope. The models of \citet{Dessart2016} currently implement two mixing schemes. Thus, stronger or more directed mixing, such as that described in \citet{Hammer2010}, may be required. However, while such effects may reconcile observations of some Type IIb and Ib SNe, it is unclear if they can similarly explain the trends observed in Type Ic and Ic-BL SNe---which also display low $\beta$ values, but do not have any detectable He in their spectra. While the presence of He in the progenitors of Type Ic SN is still debated \citep[e.g.][]{Hachinger2012}, arguments rely on the He being transparent. Mixing of $^{56}$Ni into a He envelope would significantly increase the likelihood of non-thermal excitation and thus observed spectroscopic features, making this explanation less plausible \citep{Dessart2012}. 

\subsubsection{Mixing of Radioactive Material}

The mixing of radioactive material within SN ejecta is generally difficult to model due to its inherent 3D nature \citep[e.g.][]{Joggerst2009,Hammer2010,Wongwathanarat2015}. While it is generally believed that the \nickel distribution of SESN is more centrally concentrated than in Type Ia SN, there is also evidence that at least some mixing is required to reproduce SESNe observations \citep{Dessart2012}. The distribution of \nickel inside the ejecta can alter the shape of the light curve and thus impact the $\beta$ parameter. In particular, for smoothly stratified models, more extended \nickel distributions (corresponding to stronger mixing) will lead to both shorter rise times and higher luminosities for a given \Mni\ (e.g., \citealt{Dessart2016}, KK19). However, KK19 find that these two effects combine to produce a \emph{higher} $\beta$ value for more strongly mixed models. They find that the lowest $\beta$ that can be achieved for a constant opacity model is $\beta$ $=$ 4/3 for a centrally concentrated heating source. Thus, while mixing of radioactive elements can modify the rise time and luminosity of SNe, it is likely that these would need to be coupled with the composition and opacity effects described in \S~\ref{sub:composition} to explain the low $\beta$ values of the observed SESNe. 

{We emphasize that these conclusions are applicable to mixing processes that lead to smoothly stratified (1D) \nickel distributions. Future theoretical work will be needed to fully assess the impact of larger scale mixing processes due to 3D explosions (e.g.\ \citealt{Couch2015}). We acknowledge that the impact of this uncertainty may explain the discrepancy between the $\beta$ of the observed SESNe and numerical models, highlighting the need for further work. }

\subsubsection{Asymmetry}
There is growing evidence from a combination of spectropolarimetry, nebular spectroscopy, and resolved SN remnants that some SESN may be asymmetric \citep[e.g.][]{Valenti2011,Milisavljevic2015,Tanaka2017}. As shown in Figures \ref{fig:models} and \ref{fig:modelparam}, the $\beta$ values of the mildly asymmetric simulations of \cite{Barnes2018} depend on the observer's viewing angle. This is primarily due to a variations in the observed peak luminosity, with viewing angles with larger $L_{\rm p}$ leading to lower $\beta$ values. Thus, depending on their nature, asymmetries in \nickel mixing or ejecta distribution can be reflected both in the mean value and scatter in the $\beta$ parameters observed for a population of SESNe.

\begin{figure}
    \centering
    \includegraphics[width=0.5\textwidth]{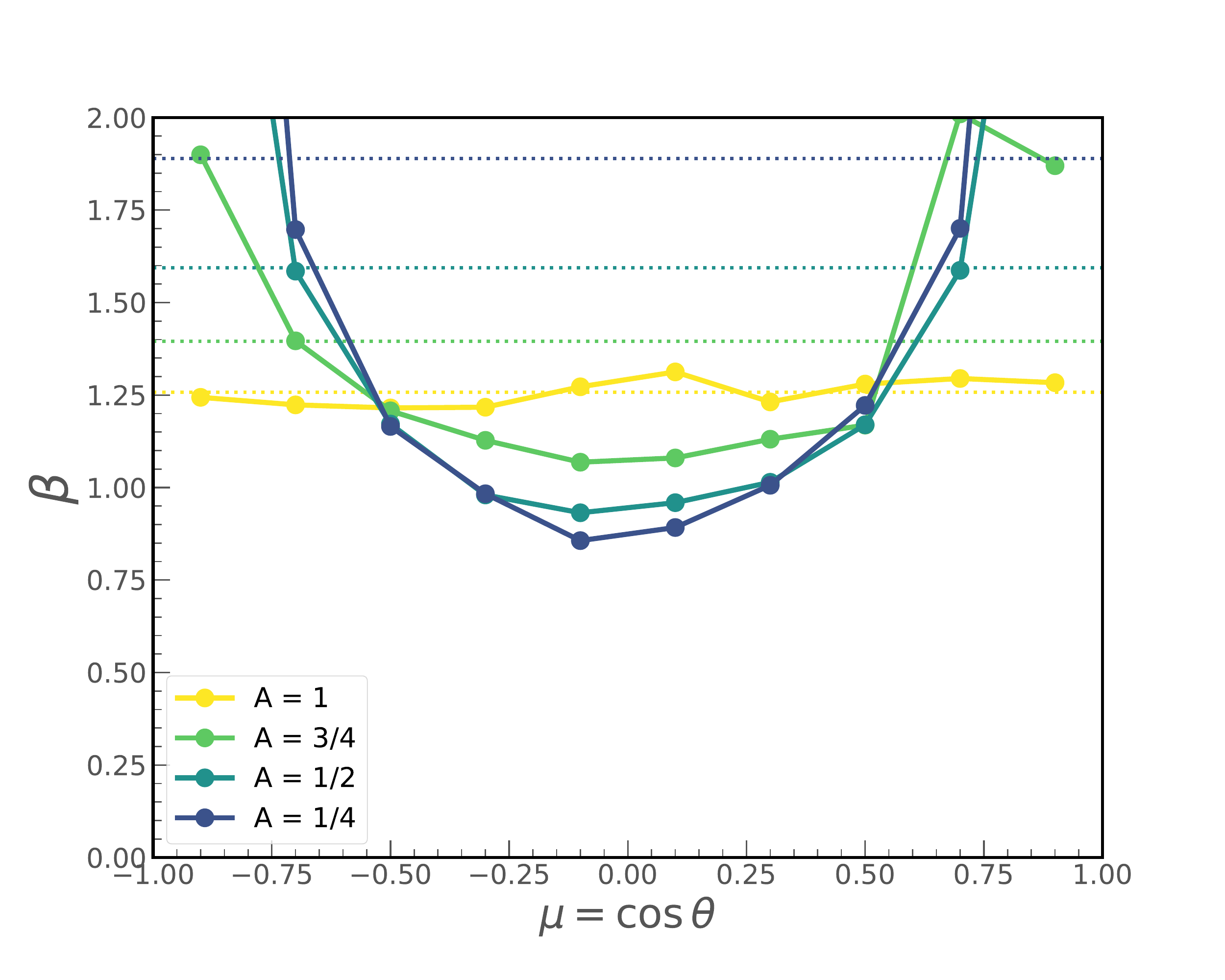}
    \caption{Inferred value of $\beta$ for the set of 2D radiative transfer models, depending on viewing angle $\mu=\cos\theta$, where $\mu=\pm 1$ corresponds to the pole and $\mu=0$ corresponds to the equator.  Different lines correspond to model ejecta with varying axis ratios $A=v_r/v_z$ (i.e. degree of asymmetry), where $A=1$ (yellow line) corresponds to a spherically symmetric ejecta. Dashed horizontal lines indicate the inferred $\beta$ when averaged over all viewing angles. We find that while line-of-site effects can lead to significant scatter in observed $\beta$, asymmetry will \emph{increase} the average $\beta$ found for a population.}
    \label{fig:beta_asymmetry}
\end{figure}

To test the degree to which asymmetry can modify observed $\beta$ values for a population, we run a set of light curve simulations with varying degrees of ejecta asymmetry using the multi-dimensional radiative transfer code Sedona \citep{Kasen2006}. We assume an axisymmetric homologously expanding ejecta profile consisting of a broken power-law in density following \cite{Chevalier1989} and \cite{Kasen2016}. To account for deviations from spherical symmetry, we vary the semi-major axis as in \cite{Darbha2020} paramterized by $A\equiv v_r/v_z$, where $v_r$ and $v_z$ are the outer ejecta velocities at the equator and pole, respectively. In total, we run four radiative transfer simulations with $A\leq 1$, i.e., prolate ejecta configurations. 

We choose fiducial values of $M_{\rm ej}=2~M_\odot$, $v_{\rm z}=10^4$ km s$^{-1}$ for our simulations. To account for the heating, we set the innermost $0.1~M_\odot$ to consist of $^{56}$Ni. Finally, we assume a constant grey opacity of $\kappa=0.2$ cm$^2$ g$^{-1}$. The resulting light curve is then calculated at ten different viewing angles $\mu=\cos\theta$, in the range $\mu\in\{-1,1\}$, where $\mu=0$ and $\mu=\pm 1$ view the ejecta along the equator and poles, respectively.

We measure the different values of $L_{\rm p}(\mu,A)$ and $t_{\rm p}(\mu,A)$ for the output bolometric light curves, which we then map onto an inferred $\beta$ based on Equation~\ref{eq:khatami} and the model parameter $M_{\rm Ni}=0.1 ~M_\odot$. In Figure~\ref{fig:beta_asymmetry}, we show the inferred values of $\beta$ for the different set of asymmetric ejecta configurations and viewing angles. As expected, $\beta$ does not vary with viewing angle for the case $A=1$, i.e. no ejecta asymmetry. However, increasing the degree of asymmetry results in lower inferred $\beta$ when viewed along the equator, and higher $\beta$ when viewed at the poles. This is due to $L_{\rm p}$ being larger when viewed along the equator, where the projected surface area is largest; similarly, $L_{\rm p}$ is decreased along the poles for asymmetric ejecta due to a smaller projected surface area \citep{Darbha2020}. This is in agreement with the results found in \cite{Barnes2018}.

While the total spread in $\beta$ values observed for our astymmetric simulations is $\gtrsim$1---well matched to the scatter in our observed population---when averaged over all viewing angles, we find that \textit{asymmetry acts to increase the mean inferred value of} $\beta$. This is opposite to the direction that has been observed, where the average $\beta$ is systematically \textit{lower} than the spherically symmetric models of \cite{Dessart2016,Ertl2019} (as well as the $A=1$ model run in this work). Thus, insofar as asymmetry can be represented by an expanding broken power-law ellipsoid, we conclude that asymmetry, although possibly explaining some of the scatter seen in Figure~\ref{fig:models}, cannot explain the \textit{systematically} lower inferred values of $\beta$. However, we note that when asymmetry is strong other effects such as the development of non-radial flows \citep{Matzner2013,Afsariardchi2018} and the ejection of nickel-rich clumps to high velocities \citep{Drout2016} could further influence the light curve morphology.

\subsubsection{Additional Power Sources}
\label{sub:additional_power}

In \S~\ref{sub:lowbeta} we demonstrated that the main driver of the low $\beta$ values for our observed SESN was their high $L_{\rm p}$ for a given \Mni\ (Figure~\ref{fig:modelparam}). Thus, another plausible explanation for the origin of the discrepancy between the $\beta$ distribution of the models and the observed SESNe is that additional power sources beyond the radioactive decay of $^{56}$Ni contribute to the peak luminosity of SESNe. This was previously proposed by \citet{Ertl2019}, who noted that their numerical simulations were unable to reproduce the brighter half of observed Type Ib/c SNe luminosity function (Figure~\ref{fig:modelparam}).  In addition, when modeling a sample of SESN with the luminosity integral method of \citet{Katz2013}, \citet{Sharon2020} required an additional model parameter, which they interpret as non-negligible amount of emission produced by power sources beyond $^{56}$Ni.  The presence of and additional luminosity source near peak could also explain why the observed SESNe show an even larger discrepancy between Arnett and tail-measured \Mni\ than the theoretical models of \citet{Dessart2016}.

Within our sample, the need for additional power sources is particularly conceivable for SN\,1994I and SN\,2007ru, for which we derived a negative and close to zero $\beta$, respectively. In practice, a negative $\beta$ is not physical in the general definition given by KK19 (see Equation~\ref{eq:general}). Rather, this implies that the version of this equation which assumes pure \nickel heating (Equation~\ref{eq:khatami}) is \emph{incapable} of producing such a bright peak luminosity on the observed rise time when coupled with the \Mni\ measured from the radioactive tail. While these are the most extreme cases, other observed SESN may also have extra power sources contributing to the observed luminosity. In this case, attributing the observed peak luminosity solely to the radioactive decay of \nickel will result in smaller $\beta$ values that expected based on their composition and \nickel distribution. 

Besides the radioactive decay of \nickelwospace, power sources that can contribute to the light curves of CCSN include shock cooling emission, ejecta interaction with circumstellar material (CSM), and energy from a central engine. Here, we assess the viability of each of these sources in explaining the discrepancy between observed and model SESNe, and implications thereof.

{\bf \emph{Luminosity Required:}} We begin by evaluating how much excess luminosity would be required to reconcile the $\beta$ values we derive in \S~\ref{sub:khatami} with those of the theoretical models in \S~\ref{sub:diffmodels}. To do so, we assume that the average value of $\beta$=1.125 found by the models of \citet{Dessart2016} is accurate for SESN powered only by radioactive decay. Given their full treatment of radiative transfer/opacity, this is equivalent to assuming that the composition, energetics, and mixing of radioactive material included in \citet{Dessart2016} reflect reality. Then, using $\beta = 1.125$, the \Mni\ and $t_{\rm p}$ listed in Table~\ref{tab:sn_derived_param}, and Equation~\ref{eq:khatami}, we calculate the luminosity that \nickel decay can produce at the peak time of the SN. From this, we define an ``excess luminosity factor'', $f$, as the fraction of peak luminosity that is in excess over what is expected from a $\beta=1.125$ explosion with a nickel mass as defined by the light curve tail. 

\begin{figure}[t!]
\centering
\includegraphics[trim=0.1cm 0.7cm  0.2cm 1cm, clip=true, scale=0.44]{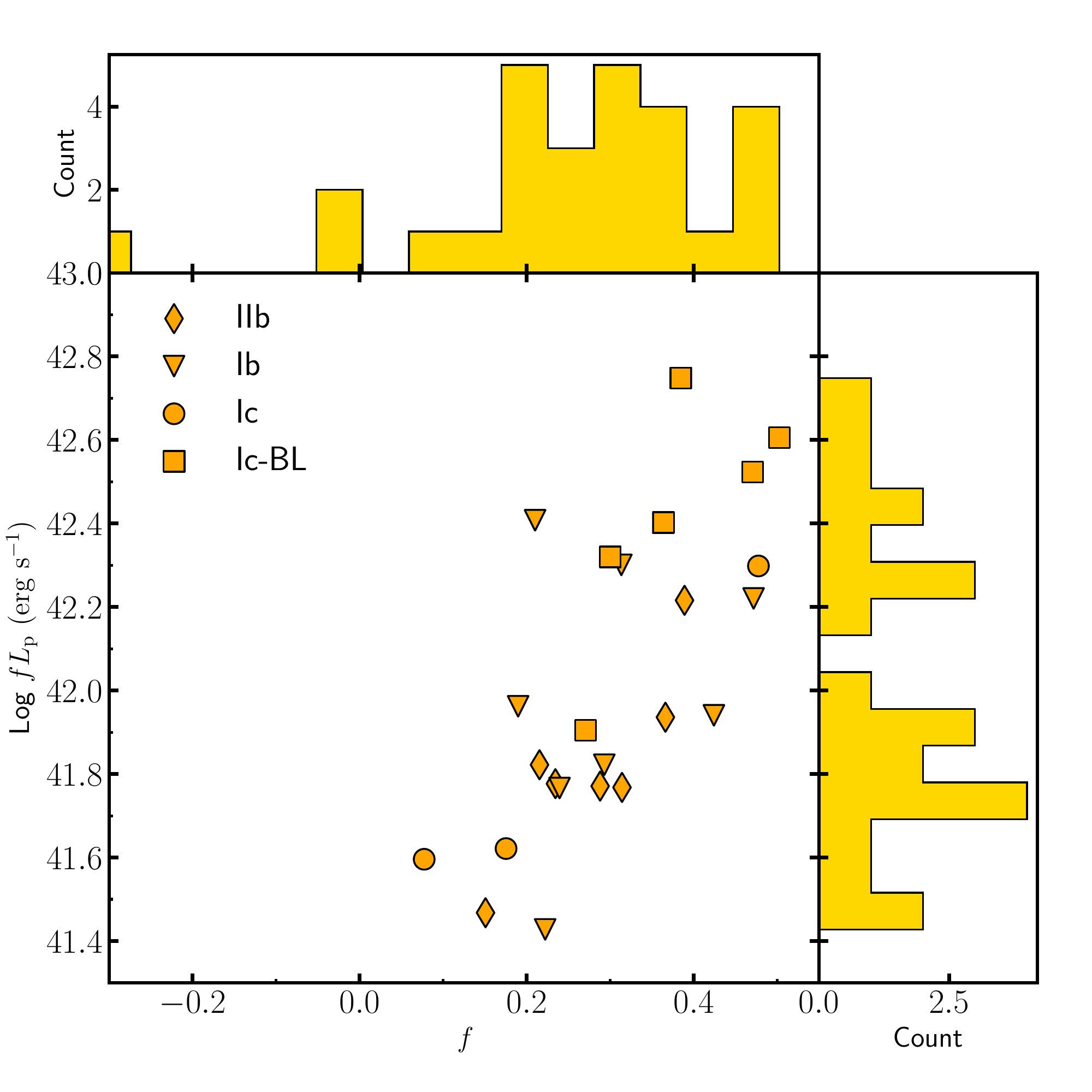}
\caption{Excess power factor, $f$, versus logarithm of excess peak luminosity $f L_{\rm p}$ for our SESN sample. Histograms of $f$ and $f L_{\rm p}$ are displayed on x- and y-axis, respectively. Overall we find that reconciling the observed SESNe with current numerical simulations would require 7--50\% of their peak luminosity to come from power sources other than \nickelwospace. This corresponds to excess luminosities of 41.4 $\lesssim$ $\log (f L_{\rm p})$ $\lesssim$ 42.8.}
\label{fig:extra_lum}
\end{figure}

The resulting values of $f$ for each SN are listed in Table~\ref{tab:sn_derived_param} and the mean, median and standard deviation in Table~\ref{tab:beta_stat}. In Figure~\ref{fig:extra_lum} we plot the excess luminosity factor, $f$, versus the excess luminosity, $f \times L_{\rm p}$. We emphasize that these ``excess luminosity'' values should be taken as order of magnitude estimates only, as they (a) neglect any variations from $\beta=1.125$ that can be caused by effects such as enhanced nickel mixing and asymmetry and (b) inherently assume that the radioactive component peaks at the same time as the observed light curve; depending on the nature of the excess luminosity source this need not be the case. Never-the-less, they provide a useful diagnostic.

Overall, we find $f$ values in the range of $-0.33$ to $0.5$. Three events (SN\,1996cb, SN2009bb, and SN\,2010bm) have negative $f$ values, reflecting the fact that they had measured $\beta$ $>1.125$, and thus do not \emph{require} additional power sources. For the remaining 24 events, we find that if the models of \citet{Dessart2016} and our tail \Mni\ are accurate, we require that $\sim$7--50\% of the their peak luminosities come from sources other than \nickelwospace. This translates into considerable excess luminosity in the range of 2.5$\times$10$^{41}$ -- 5.5$\times$10$^{42}$ erg s$^{-1}$ or, equivalently, peak bolometric magnitudes for the excess power source of $-$14.9 mag $>$ M$_{\rm{bol}}$ $>$ $-$18.2 mag. Out of the 6 events that would require the most excess luminosity, five are Type Ic-BL, while the events which require the highest \emph{fraction} of the peak luminosity to come from other sources, are mixed between sub-types.

{\bf \emph{Shock Cooling and CSM Interaction:}} Both shock cooling and CSM interaction face challenges in explaining this required excess emission. First, shock cooling emission is predicted to be both faint and short-lived for the compact progenitors (R $\sim$ R$_\odot$) typically evoked for H-poor SESN \citep{Nakar2010,Rabinak2011}. Second, the substantial luminosity required and lack of narrow emission lines in the majority of SESNe spectra limit any potential CSM to dense and confined shell or disk-like configurations \citep{Chevalier2011,Moriya2012,Smith2015}, which are not predicted from standard models of stellar evolution \citep{Smith2014}. However, recent progress may change this picture: models show that particularly low mass He stars ($\lesssim$3 M$_\odot$) can undergo substantial inflation (achieving radii $\gtrsim$100 R$_\odot$; e.g.\ \citealt{Yoon2010}, \citealt{Kleiser2018b}, \citealt{Woosley2019}) and there is increasing observational evidence that many CCSN---of all varieties---undergo a period of enhanced mass loss shortly before core-collapse \citep[e.g][]{Margutti2015,Drout2016,Bruch2020}. Thus, a significant fraction of SESN progenitors may have large effective radii ($\gtrsim$ 20 R$_\odot$) in which case shock deposited energy can contribute substantially to their observed luminosity. 

As described in \S~\ref{sub:specialized}, \citet{Kleiser2018b,Kleiser2018a} model the light curves that would result solely from the diffusion of shock deposited energy in both of these scenarios.
The set of O/He-rich CSM shells modeled in \citet{Kleiser2018a} have masses of 1$-$4 M$_\odot$ and are located at radii of $\sim$15-60 R$_\odot$, designed to mimic an intense final mass-loss episode. They find transients which rise to peak magnitudes of $-15~\rm{mag} < M_r < -18~\rm{mag}$ on timescales of $\sim$8--25 days---parameters which are exceptionally well matched to our estimates above for the excess luminosity required. Indeed, as shown in Figure~\ref{fig:models}, when a small amount of \nickel is added, the models of \citet{Kleiser2018b,Kleiser2018a} are able to reproduce the low $\beta$ values of the observed population. 

While the particular theoretical models plotted in Figure~\ref{fig:models} show slight double-peaked morphology in the their \emph{optical} light curves---which are not typically observed---they were created by adding \nickel to a shock cooling curve that reaches $-17$ mag and contributes $\gtrsim$50\% of the luminosity near peak. In contrast, we expect the SESN in our sample to still be dominated by the radioactive decay of \nickelwospace, with a median excess luminosity factor of $f=0.29$. In addition, a number of Type Ib/c SNe, show evidence for multiple emission components when observed early in the u-band/UV, with SN\,2013ge also exhibiting \emph{narrow} high-velocity absorption features, which may be indicative of a shell-like CSM structure \citep{Drout2016,Kleiser2018a}. However, the lack of \emph{prominent} double-peaked structure in many Type Ib/c SNe combined with the need for excess luminosity near maximum light implies that: (i) the He/O-rich material which leads to the large effective radius cannot be too tenuous, in which case it would become optically thin on a timescale of $\sim$days \citep{Dessart2018,Woosley2019} and (ii) the \nickel synthesized in the explosion must be well-mixed to avoid a significant dark-period and subsequently large offset in time between the two emission components \citep{Kleiser2018a}. 

We therefore conclude that shock cooling emission is a viable source for the excess luminosity required to reconcile SESN observations and models. While low mass He stars can reach the radii required through ``natural'' stellar evolution processes, such events are predicted to eject very little radioactive \nickel \citep{Kleiser2018b,Woosley2019}. Thus, reproducing the full observed population via this mechanism likely requires a significant number of SESN to undergo intense late-stage mass loss due to instabilities \citep{SmithArnett2014,Woosley2019}, internal gravity waves \citep{Quataert2012,Fuller2018}, or some other physical process during the final nuclear burning stages. We emphasize that this shock cooling emission will rapidly fade once the ejecta cools to the recombination temperature of O/He-rich material ($t < 40$ days), and thus \Mni\ measured at $>$60 days post-explosion should be unaffected.

{\bf \emph{Central Engines:}} On the other hand, a central engine could provide the additional energy source required to power the light curve around peak. We note, in particular, that the we found a lower mean $\beta$ value and require a higher mean excess luminosity for Type Ic-BL SNe (Table~\ref{tab:beta_stat}), for which central engines are commonly evoked. Both magnetars and collapsars can provide a natural source of extra luminosity, in the form of spin-down energy and fallback accretion \citep{Dexter2013}, respectively. \citet{Ertl2019} investigate the former as a power source for SESNe in general, arguing that even the formation of a more moderate millisecond pulsar (e.g., \citealt{Yoon2010}) may be sufficient for rotational energy to impact the SN light curve without impacting the overall energetics.

\begin{figure}[t!]
\centering
\includegraphics[trim=0.2cm 0.2cm  0.2cm 0.2cm, clip=true, scale=0.55]{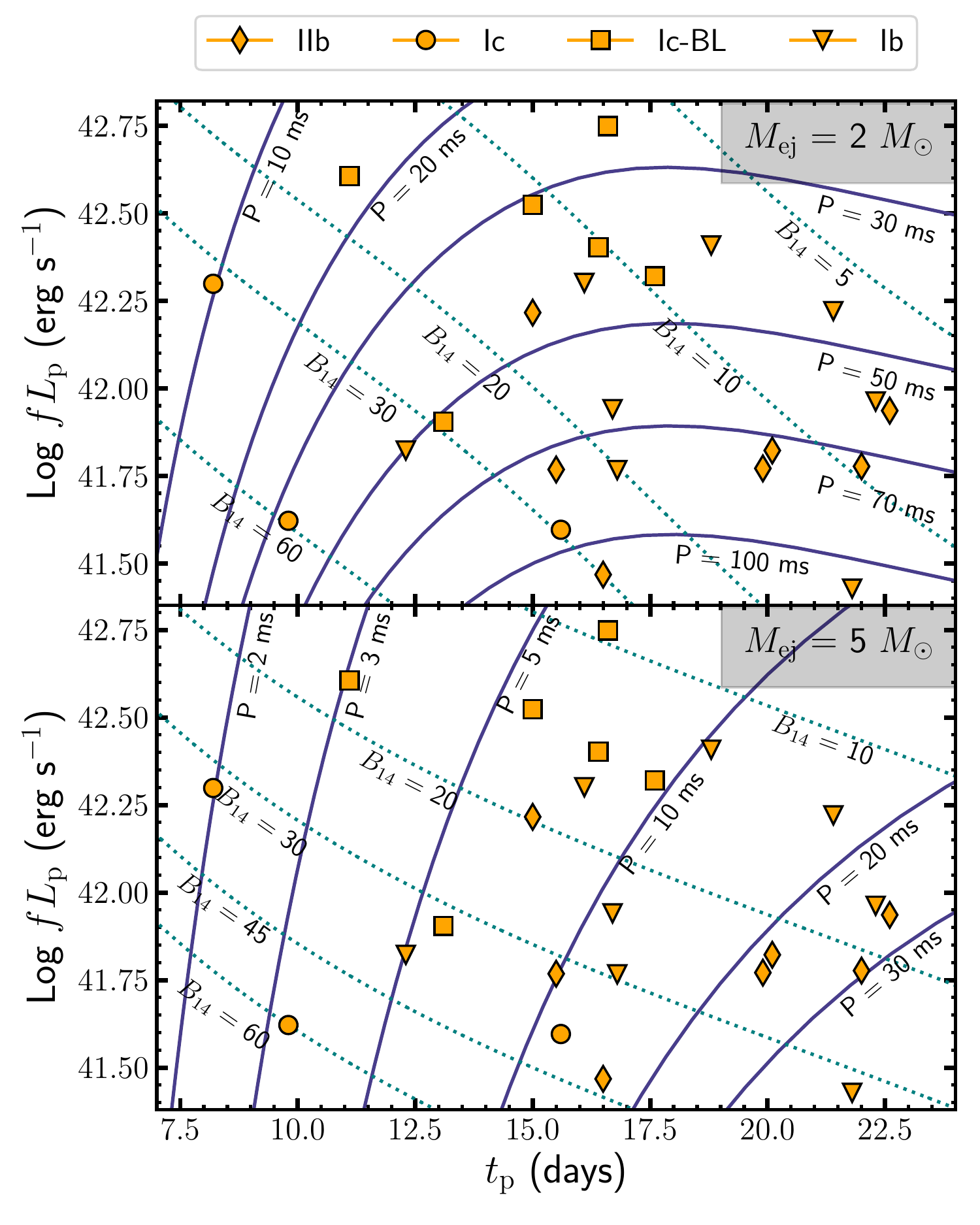}
\caption{Logarithm of excess peak luminosity $\log (fL_{\rm p})$ versus peak time $t_{\rm p}$ for our sample of SESNe (excluding SNe with negative excess factor $f$). The diamond, inverted triangle, circle, and square orange markers represent SN types IIb, Ib, Ic, and Ic-BL, respectively. Also plotted are lines which which represent the peak luminosity and time achieved by magnetar spin-down models with a fixed period but varying magnetic field strength (solid purple) and fixed magnetic field but varying periods (dotted cyan) for ejecta mass of 2~\Msolar\ (top panel) and 5~\Msolar\ (bottom panel).}
\label{fig:magnetars_phasespace}
\end{figure}

We use the magnetar spin-down energy and timescale of \citet{Kasen2017} together with SN timescale and ejecta velocity of \citet{Kasen2016} to estimate the general phase space of magnetar properties required to provide the excess luminosity found above. In Figure~\ref{fig:magnetars_phasespace} we plot rise time vs.\ excess luminosity for our sample of SESNe. Also shown are lines which which represent the peak luminosity and time achieved by magnetar models with a fixed period but varying magnetic field (solid) and fixed magnetic field but varying periods (dotted). All models assume an opacity of $\kappa=0.1$~cm$^2$~g$^{-1}$, explosion energy of $10^{51}$~erg, and were calculated both for $M_{\rm ej} = 2$~\Msolar\ (top panel) and $M_{\rm ej} = 5$~\Msolar\  (bottom panel)---chosen to span the range of SESNe ejecta masses from \citet{Lyman2016}. For $M_{\rm ej}=2$~\Msolar, $P$ values of 10--116 ms and $B_{14}$ ($=B/10^{14}$~G) values of 7--59 are yield luminosities consistent with requirements. For $M_{\rm ej}=5$~\Msolar, similar magnetic field strengths but shorter periods are necessary with $P =$ 2--32 ms and B$_{14} =$ 10--60 spanning the phase space occupied by most SESNe in our sample. The ranges of $P$ and $B_{14}$ found here overlap with---but extend to longer periods and higher magnetic field strengths---than those of magnetar fits to super-luminous SN light curves by \citet{Nicholl2017}.

\begin{figure}[t!]
\centering
\includegraphics[trim=0.5cm 0.6cm  0.5cm 0.5cm, clip=true, scale=0.63]{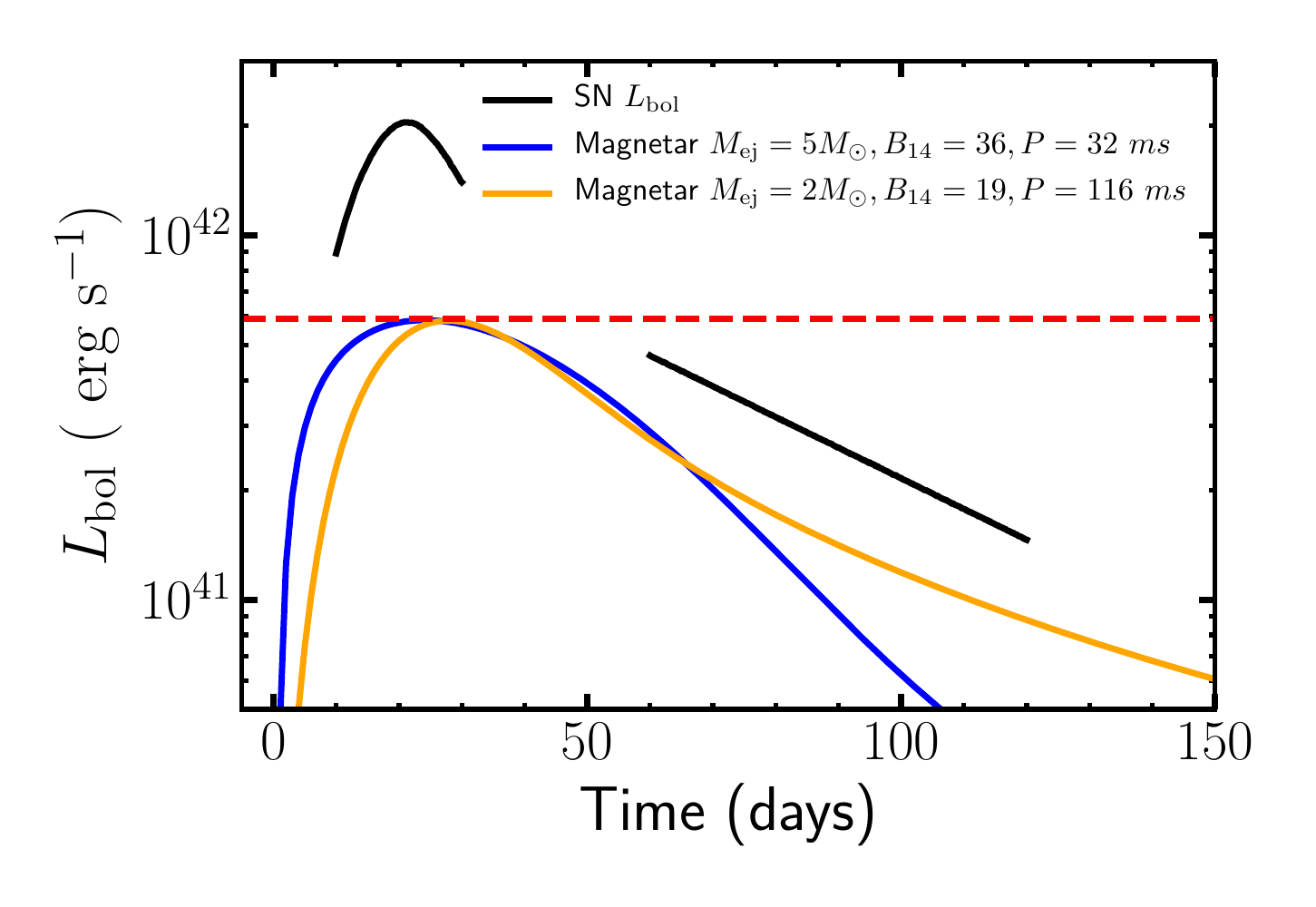}
\caption{The bolometric light curve of SN2008ax in comparison with two magnetar models. The dashed red horizontal line indicates the excess emission in the $L_{\rm p}$ of SN2008ax (i.e, $\sim$29\% of the peak luminosity). If this power is contributed to the light curve by sources other than the radioactive decay, then $\beta$ increases to the same level as predicted by numerical models, i.e., $\beta \simeq 1.125$, shown in Figure~\ref{fig:models}. The orange and blue curves represent the light curves of two magnetar models with the peak time and luminosity similar to the $t_{\rm p}$ and $0.29 \times L_{\rm p}$ of SN2008ax. }
\label{fig:magnetars_lightcurve}
\end{figure}

Figure~\ref{fig:magnetars_lightcurve} displays two representative magnetar models for the excess peak luminosity of SN2008ax. The excess emission factor is 0.29, corresponding to a luminosity of $5.90 \times 10^{41}$~erg s$^{-1}$ (dashed line). The blue curve illustrates a model with $B_{14} = 36$, $P=32$~ms, and \Mej~=~5~\Msolar, while the orange curve represents a magnetar with $B_{14} = 19$, $P=116$~ms, and \Mej~=~2~\Msolar. Both models peak at the level of the excess luminosity calculated above, but have have different magnetar properties and light curve morphologies.  Figure~\ref{fig:magnetars_lightcurve} highlights that, unlike shock cooling emission, if a magnetar contributes to the light curve near peak it can also power a non-negligible portion of the tail luminosity. In this case, tail-based measurement of \Mni\ would be overestimated and, as a result, the fraction of the peak luminosity which must come from sources other than radioactive decay \emph{underestimated}. While we find that fits over 60--120 days post-explosion, as performed in \S~\ref{sec:lbol}, are not sufficient to distinguish between the $L(t) \propto t^{-2}$ power-law decline of the magnetar model and the exponential form of the radioactive decay of \nickel (Equation~\ref{eq:lni}) future modelling of a subset of SESNe with data covering a longer baseline ($\gtrsim$200 days) could potentially break this degeneracy. If the slope of the tail is shallower than $^{56}\text{Co} \rightarrow ^{56}\text{Fe}$ conversion rate this could also be an indication that other power sources are contributing to the light curve tail \citep[e.g.,][]{Ertl2019}. However, none of the SESNe in our sample have such a shallow tail.

\subsubsection{Possible Limitations of Tail \Mni\ Values}
\label{sub:mnilimit}

Throughout the analyses of \S~\ref{sec:res} and \S~\ref{sec:disc}, and specifically in the model comparisons of Figures~\ref{fig:models} and~\ref{fig:modelparam}, we assume that the \Mni\ values measured from radioactive decay modeling of SESNe light curve tails are reliable estimates for the actual \Mni. In \S~\ref{sec:nickel} we utilize the tail luminosity model of \cite{Wygoda2019}, which relies on a number of assumptions. In particular, the simple scaling form of the $\gamma$-ray deposition factor, $f_{\rm dep}=1-e^{-(t/T_0)^{-2}}$, only holds if the ejecta is in homologous expansion and if the $\gamma$-ray opacity is constant and purely absorptive. However, both are standard assumptions: ejecta should reach a phase of homologous expansion after several expansion doubling times---i.e., a few days---while we measure \Mni\ over epoches $>$60 days, and the $\gamma$-ray opacity is typically assumed to be constant \citep{Sutherland1984, Clocchiatti1997, Wygoda2019}. We note that other assumptions on the ejecta density distribution or nickel mixing will chiefly change pre-factors of $T_0$, while the scaling relation of $f_{\rm dep}$---and therefore \Mni---will remain unchanged. 

Thus, we conclude that any error in \Mni\ is due to the late-time \emph{luminosity} that we attribute to radioactive decay not being accurate. This could occur if either the bolometric corrections utilized in \S~\ref{sec:lbol} do not adequately describe the behavior of our observed sample or if power sources other than radioactive decay (e.g., CSM interaction, magnetar spin down; see above) contribute to the luminosity on the light curve tail. In the latter case, our tail \Mni\ values would be overestimated. However, we emphasize that this effect can not resolve the tension with theoretical models show in Figures~\ref{fig:models} and~\ref{fig:modelparam}. Rather, lower \Mni\ values would \emph{increase} the discrepancy, with the observed sample of SESNe showing even larger $L_{\rm p}$ values for a given \Mni.

\subsection{Consequences of \Mni\ Discrepancy with Type II SNe}
The discrepancy in \nickel mass distributions for SESNe and Type II SN was first identified using Arnett-based measurements of \Mni\ for SESNe \citep{Anderson2019}. In \S~\ref{sub:comp} we demonstrated that Arnett's rule over-predicts \Mni\ for SESNe by roughly a factor of 2---likely due \emph{both} to limitations in Arnett's model (KK19) and to possible contributions from additional power sources to the peak luminosity of SESN (\S~\ref{sub:additional_power}). However, in \S~\ref{sub:compTypIIvsSESNe} we find that even when tail-based \Mni\ values are used, our population of SESNe have a mean \Mni\ value (0.12~\Msolar) which is a factor of $\sim$3 higher than Type II SNe (0.044~\Msolar). Therefore a critical question remains: what makes the \Mni\ distribution of SESNe skew to larger values than Type II SNe? Are the \Mni\ values for the \emph{observed} population of SESNe biased? If not, does the discrepancy with Type II SNe come about because SENe originate from higher ZAMS mass stars? Or are the ZAMS masses of SESNe and H-rich Type II SNe are somewhat similar, but other physical mechanisms are responsible for the observed difference in the \Mni\ distribution? Here, we consider each of these questions in turn.

\emph{Does the distribution of \Mni\ derived accurately represent the true distribution of \Mni\ synthesized in the core-collapse of H-poor stars?} As highlighted by \citet{Meza2020}, the discrepancy between SESNe and Type II SNe is primarily due to a \emph{lack} of observed SESNe with low \Mni. While the light curves of Type II SNe---powered predominately by H recombination---remain luminous for $\gtrsim$100 days regardless of \Mni, low-\Mni\ SESN may be faint and/or rapidly-evolving. In particular, as described in \S~\ref{sub:additional_power}, low-mass stripped He stars (M$\lesssim$3--4 M$_\odot$; corresponding to ZAMS $\lesssim$12--15 M$_\odot$) can inflate to large radii prior to core-collapse. Stars with these initial masses likely dominate the population of Type II SNe, due to the initial mass function. However, the explosion of their stripped counterparts may be dominated by bright and rapidly-fading cooling envelope emission \citep{Dessart2018,Kleiser2018b,Woosley2019}, with minimal contributions from \nickelwospace. Observationally, these could manifest as rapidly-fading Type I SN \citep{Kasliwal2010,Drout2013}, Type Ibn SNe \citep{Hosseinzadeh2017}, or the broader class of fast blue optical transients \citep{Drout2014} rather than ``traditional'' SESNe. Such events were not explicitly included in our sample, and have rarely been followed to late enough times to constrain the \Mni\ ejected. While the rates of rapidly-evolving transients ($\sim$4-7\% of the CCSN rate; \citealt{Drout2014}) are not sufficient to fully resolve the discrepancy, they could reduce its significance.

Alternatively, the \Mni\ distribution shown in Figure~\ref{fig:anderson_cdf} may not accurately represent reality if the light curves of SESNe are not solely powered by the radioactive decay of \nickel and the timescale of the additional power source(s) is $\gtrsim$60 days (see \S~\ref{sub:additional_power}). In this case, attributing the full tail luminosity to \nickel would lead to an overestimate of the true values. Within this context, it is notable that while simulations of neutrino-driven core-collapse SNe \citep{Sukhbold2016,Ertl2019} are able to self-consistently produce the range of \Mni\ values observed in H-rich Type II SN ($\sim$0.004--0.13~\Msolar; \citealt{Muller2017}, \citealt{Afsariardchi2019}), they are unable to produce \Mni\ values as high as those derived for the upper $\sim$30\% of our SESN sample (see Figures~\ref{fig:models} and~\ref{fig:modelparam}). It was for this reason that \citet{Ertl2019} invoked magnetars to explain the observed light curves of SESNe. While this may resolve the \Mni\ discrepancy, it subsequently requires that magnetars influence SESN at a higher rate than Type II SNe. This may be a natural consequence of stripped stars retaining a larger fraction of their angular momentum, which in H-rich stars is lost primarily due to rotational breaking when the star expands to the RSG phase \citep[e.g.,][]{Ertl2019}.

\emph{Do SESNe originate from higher ZAMS mass stars than Type II SNe?} If observational biases cannot explain the relative lack of low \Mni\ SESNe, the discrepancy could indicate that SESNe preferentially form from higher ZAMS mass stars---which are expected to synthesize higher amounts of \Mni---than Type II SNe. While on face value this may favor the single star progenitor channel, this is in tension with the observed occurrence rate of SESNe, which is much higher than the explosion rate of stars with ZAMS mass $\gtrsim25$~\Msolar\ that eventually become WR stars \citep{Smith2011}. In addition, the mass-loss rate of massive stars is still uncertain, and it is not clear whether all massive stars with ZAMS mass $\gtrsim25$~\Msolar\ can strip their H envelope \citep{Smith2014}. Another possible explanation is that SESNe do form stripped binaries and span the same \emph{overall} ZAMS masses as Type II SNe. However, while the relative rates of Type II SNe are dominated by the IMF, SESNe are skewed to higher relative masses. This may be a natural consequence of the close binary fraction being larger for high mass stars \citep{Moe2017,Moe2019}, although detailed population synthesis calculations are necessary to test this hypothesis. We note that the population of SESNe coming from a distribution skewed to higher ZAMS masses, and hence shorter lifetimes, would be consistent with the fact that they are found closer, on average, to sites of active star formation than Type II SNe \citep{Anderson2012}.

\emph{Do additional physical mechanisms modify the \Mni\ synthesized in SESN?} Alternatively, if the \Mni\ distribution for SESNe is accurate, and SESNe originate from similar ZAMS masses as H-rich Type II SNe, then there must be some physical processes that make the \Mni\ of SESNe larger. While it is often assumed stripping of the H and even He envelope via binary interaction should leave the inner core structure of the primary star intact \citep[e.g.,][]{Fryer2001}---in which case the \Mni\ distributions of SESN and Type II SNe should be indistinguishable---this picture may not be complete. In particular:

\begin{itemize}
    \item[(i)] In close binaries, fast orbital rotation, tides, magnetic breaking, and angular momentum transport can influence the convective core sizes profoundly \citep[e.g.,][]{Song2018}. Therefore, changes in the core structure may impact the \Mni\ production.
    
    \item[(ii)] {\nilou A fraction of SESNe may originate from the merger of binary stars \citep{Zapartas2017}. In this case, a more massive core capable of producing a larger amount of \nickel may result. Although the rate of this merger channel seems to be relatively small \cite[$\sim12\%$ of SESNe;][]{Zapartas2017}, it will boost the overall \Mni\ of SESNe produced via the binary channel. }
    
\end{itemize}

Additional analysis is required to distinguish the contributions of each of the above scenarios towards explaining the discrepancy in observed \Mni\ for H-poor and H-rich core-collapse SNe.

\section{Summary and Conclusions} \label{sec:conc}

In this paper we measure the \nickel masses for a sample of 27 stripped-envelope core-collapse SNe with well-constrained explosion epochs and late-time photometric coverage by modeling their radioactive tails. We both compare these results to Arnett-based \Mni\ measurements and use them, in conjunction with the observed rise times and peak luminosities for the sample, to \emph{observationally} calibrate the $\beta$ parameter in the new analytic light curve model of \citet{Khatami2019}. This parameter $\beta$ allows the internal energy of the ejecta to lag or lead the observed luminosity at the time of peak (in contrast to Arnett models), and is hence a function of the ejecta composition, mixing, asymmetry, and total power sources. Here we summarize our main conclusions. 

\begin{enumerate}
    \item We find \nickel masses for measured from the radioactive tail of 0.03 \Msolar\ $<$ \Mni\ $<$ 0.57 \Msolar, with a median value of 0.08 \Msolar. Type Ic-BL SNe show higher \Mni\ on average, with a median value of 0.15 \Msolar.
    \vspace{-0.08in}
    \item \Mni\ values measured via Arnett's rule are systematically larger than those found from the radioactive tail by a factor of $\sim$2. While limitations in Arnett's rule when applied to SESNe had previously been discussed, this discrepancy is approximately a factor of 2 larger than that found in recent numerical simulations \citep{Dessart2016}.
    \vspace{-0.08in}
    \item Using our observed rise times, peak luminosities, and tail-based \Mni\ values we find KK19 $\beta$ values which range from 0.0 $<$ $\beta$ $<$ 1.71, with a median value of 0.70. The calibrated $\beta$ values show significant spread with a standard deviation of 0.34. Two objects exhibit $\beta \approx 0$, which may indicate that the radioactive decay of \nickel is incapable of powering their entire peak luminosity.
    \vspace{-0.08in}
    \item Despite the observed scatter, we demonstrate that using the model of KK19 with the median values of our calibrated $\beta$ (see Table~\ref{tab:beta_stat}) yields significantly improved measurements of \Mni\ in comparison to Arnett's rule when only photospheric data is available.
    \vspace{-0.2in}
    \item When comparing our calibrated $\beta$ values to those of inferred from a range numerical light curve models \citep[e.g.][]{Dessart2016,Ertl2019}, we find that the simulations significantly overestimate $\beta$, on average. This is primarily due to the observed sample displaying dramatically larger ($\sim$0.3--0.4 dex) peak luminosities for a given \Mni\ than the numerical models. The observed population also exhibits shorter rise times, on average.
    \vspace{-0.08in}
    \item We investigate a number of physical mechanisms to explain this observed discrepancy. Effects due to composition and the mixing of radioactive elements can lead to brighter peak luminosities and shorter rise times while the impact of line-of-site variations due to explosion asymmetries can lead the observed scatter in $\beta$. However, current models for all of these mechanisms have difficulties explaining systematically low $\beta$ values for the entire population of SESNe. Further theoretical work examining the impact of large scale mixing processes are required to assess its viability as a solution for this observed discrepancy.
    \vspace{-0.08in}
    \item Alternatively, we demonstrate that the discrepancy with numerical models can be resolved if an additional power source contributes between $\sim$7--50\% of the peak luminosity of SESNe; corresponding to luminosities in the range of 2.5$\times$10$^{41}$--5.5$\times$10$^{42}$ erg s$^{-1}$. Both diffusion of shock deposited energy and magnetar spin-down are capable of providing the required luminosity over appropriate timescales. 
    \vspace{-0.08in}
    \item Finally, we demonstrate that recently identified discrepancy between the observed \Mni\ distribution of SESNe and H-rich Type II SNe \citep{Anderson2019,Meza2020} persists in our sample. The median tail \Mni\ value of our SESNe is a factor of $\sim$3 higher than those of Type II SNe. We discuss several explanations for this discrepancy including that low \Mni\ SESN may primarily manifest as rapidly evolving transients as opposed to ``traditional'' SESN, that the close binary fraction increases for higher mass stars leading to SESN forming from a distribution of ZAMS masses skewed relative to the IMF, and that additional physical effects may impact the \Mni\ production in SESNe. 
\end{enumerate}

\acknowledgments

We thank Daniel Kasen, Katelyn Breivik, and Jennifer Hoffman for helpful comments and discussions, Tuguldur Sukhbold for providing numerical simulation results and Siva Darbha for providing code used to setup the 2D radiative transfer models.

M.R.D.\ acknowledges support from the Natural Sciences and Engineering Research Council (NSERC) of Canada through a Discovery Grant (RGPIN-2019-06186), the Canada Research Chairs Program, the Canadian Institute for Advanced Research (CIFAR), and the Dunlap Institute at the University of Toronto. This research benefited from interactions made possible by the Gordon on Betty Moore Foundation through grant GBMF5076.

D.K.K.\ is supported by the National Science Foundation Graduate Research Fellowship Program. This research used resources of the National Energy Research Scientific Computing Center, a DOE Office of Science User Facility supported by the Office of Science of the U.S. Department of Energy under Contract No. DE-AC0205CH11231.

D.S.M.\ was supported in part by a Leading  Edge Fund from the Canadian Foundation for Innovation (project No.30951) and a Discovery Grant (RGPIN-2019-06524) from the Natural Sciences and Engineering Research Council (NSERC) of Canada.

\vspace{0.7in}

\bibliographystyle{aasjournal}
\bibliography{SNeNickel}{}

\end{document}